\documentclass[
reprint,
superscriptaddress,
 amsmath,amssymb,
 aps,
 onecolumn
]{revtex4-2}
\usepackage{lineno}
\usepackage{graphicx}
\usepackage{dcolumn}
\usepackage{bm}
\usepackage{float}
\usepackage{multirow}
\usepackage{xcolor}
\begin{document}
\preprint{APS/123-QED}

\title{Internal structure of Hayward black holes}

\author{Caiying Shao}\email[E-mail: ]{shaocaiying@ucas.ac.cn}\affiliation{School of Physical Sciences, University of Chinese Academy of Sciences,Beijing 100049, China}
\author{Jun-Qi Guo}\email[E-mail: ]{sps\_guojq@ujn.edu.cn}\affiliation{School of Physics and Technology, University of Jinan, Jinan 250022, Shandong, China}
\author{Yu Tian}\email[E-mail: ]{ytian@ucas.ac.cn}\affiliation{School of Physical Sciences, University of Chinese Academy of Sciences,Beijing 100049, China}
\author{Hongbao Zhang}\email[E-mail: ]{hongbaozhang@bnu.edu.cn}\affiliation{School of Physics and Astronomy, Beijing Normal University, Beijing 100875, China}\affiliation{Key Laboratory of Multiscale Spin Physics, Ministry of Education, Beijing Normal University, Beijing 100875, China}
\date{\today}

\begin{abstract}
Regular black holes, free of central singularities, provide an ideal laboratory for probing the geometric structure of spacetime.
The global structure of some regular black holes, e.g. Hayward black hole, features an event horizon and a Cauchy horizon, raising fundamental questions about the latter's stability.
In this work, we investigate collapse of a scalar field in Hayward spacetime.
Under weak scalar perturbations, the inner horizon maintains a stable finite radius.
In the circumstance of a strong scalar field, the inner horizon shrinks to zero volume, accompanied by the formation of a spacelike singularity.
The Hayward geometry is effectively converted into a Schwarzschild-like geometry.
Furthermore, the characteristic parameter of the scalar field governs the contraction dynamics of the inner horizon.
As the parameter $p$ of the initial profile for the scalar field approaches the critical threshold ${p_*}$, the radius of the inner horizon ${r_{-}}$ exhibits a universal scaling behavior: ${r_{-}}\propto{|p - {p_*}|^\gamma}$, with a critical exponent $\gamma\approx 0.5$.
\end{abstract}

\maketitle

\section{\label{section1}Introduction}
The detection of gravitational waves from compact binary coalescences and the horizon-scale imaging of supermassive black holes have provided compelling tests of gravity in the strong-field regime~\cite{LIGOScientific:2016aoc,EventHorizonTelescope:2019dse}, firmly establishing general relativity as a robust theoretical framework for describing gravitational phenomena.
A fundamental observational challenge remains in reliably distinguishing genuine general relativistic black holes, which possess event horizons, from possible astrophysical mimickers~\cite{Murk:2022dkt}.
Furthermore, the inevitable formation of spacetime singularities in generic gravitational collapse scenarios presents profound difficulties for classical general relativity, raising fundamental questions about the completeness of the theory in extreme regimes.
It is widely anticipated that a complete theory of quantum gravity that successfully reconciles general relativity with the principles of quantum mechanics may offer a resolution to the singularity problem.
Although several promising candidates for a theory of quantum gravity have been proposed, including string theory and loop quantum gravity, these approaches are still under active development and have yet to gain consensus within the scientific community~\cite{Rovelli:1994ge}.
In light of these challenges, there is a strong motivation to investigate singularity-free black hole spacetimes within the frameworks of classical general relativity and modified gravity theories, as potential alternatives that circumvent the pathologies associated with singularities.

An effective approach to resolving the issue of singularities is to replace the central region of a black hole with a de Sitter~\cite{Ayon-Beato:1998hmi,Borde:1994ai,Hayward:2005gi} or Minkowski~\cite{Balart:2014cga,Culetu:2014lca} spacetime, thereby ensuring that the curvature invariants remain finite at the core.
Black hole solutions constructed in this manner, which avoid singularities, are collectively known as regular black holes and are regarded as alternatives to the classical black hole solutions of general relativity.
A common strategy for constructing regular black holes involves modifying the gravitational action by introducing couplings to nonlinear electrodynamic fields~\cite{Fan:2016hvf,Toshmatov:2018cks}.
The first such solution was proposed by Bardeen, who introduced a static, spherically symmetric black hole now known as the Bardeen black hole~\cite{1968qtrconf87B}.
Unlike traditional solutions that satisfy the strong energy condition and lead to singularities, these regular black holes typically violate the strong energy condition but respect the weak energy condition.
Building on these ideas, the Hayward black hole was proposed as a further refinement~\cite{Hayward:2005gi}.
Hayward’s model has inspired a wide range of regular black hole solutions, deepening our understanding on nonsingular gravitational collapse and black hole interiors~\cite{Ayon-Beato:1998hmi,Uchikata:2012zs,Balart:2014cga}.
These models are crucial for addressing foundational challenges, including Hawking’s information paradox~\cite{Hawking:1976ra,Hayward:2005gi}, and for exploring how quantum gravitational effects may regulate classical singularities~\cite{Bonanno:2000ep,Platania:2019kyx}, potentially bridging the gap between general relativity and quantum mechanics.
Due to their intriguing properties, regular black holes have become an active area of research, encompassing diverse topics such as black hole thermodynamics~\cite{Fan:2016hvf,Fan:2016rih,Guo:2021wcf}, quasinormal modes~\cite{Bronnikov:2012ch,Li:2013fka,Fernando:2012yw}, interplay between scalar fields and regular black holes~\cite{Yue:2023sep,Calza:2025yfm}, and observational features like black hole shadows~\cite{Li:2013jra,Abdujabbarov:2016hnw,Tsukamoto:2017fxq,Vagnozzi:2022moj,Pedrotti:2024znu}, among others (for a review, see e.g. \cite{Lan:2023cvz,Calza:2024fzo}).

The Hayward geometry, influenced by a nonlinear magnetic monopole, features a transition region that includes an inner horizon.
In 1968, Penrose pointed out that the inner horizons of black holes can become unstable under small perturbations, potentially leading to mass inflation, which is a phenomenon characteristic of singular black holes in general relativity~\cite{Penrose:1968ar}.
This observation raises the possibility that the inner horizons of Hayward black holes may exhibit similar instabilities.
Although the stability of inner horizons in regular black holes has been partially investigated, the question remains open~\cite{Carballo-Rubio:2022pzu,Bonanno:2023qhp}.
Building upon Ori’s model of Reissner–Nordström black holes~\cite{1991PhRvL67789O}, recent studies have extended the analysis to regular spacetimes. 
These works consider a dynamic, perturbing null shell between the event- and inner horizons to investigate its influence on the internal geometry of regular black holes~\cite{Bonanno:2020fgp,Carballo-Rubio:2021bpr}.
Nevertheless, this approach, though insightful, depends on idealized models employing approximations.
To explore more realistic physical scenarios, we consider collapse of a neutral scalar field in Hayward background, explicitly accounting for the backreaction between scalar field and geometry.
This framework allows us to explore the dynamical behavior of the Hayward geometry, particularly in the vicinity of the inner horizon and the central core.
Moreover, we examine how the inner horizon in the final state depends on the initial profile of scalar field.
Importantly, we can determine the final state of a Hayward black hole and evaluate whether a spacelike singularity may still form, potentially signaling structural breakdown.
In particular, we focus on the near-critical regime of scalar field collapse, where the radius of the inner horizon exhibits power-law scaling, and rich critical behavior emerges inside the black hole. 
This perspective offers a deeper understanding on the internal structure of Hayward black holes, revealing novel features with implications for both theoretical frameworks and observational tests.

The remainder of our paper is organized as follows.
In Section~\ref{section2}, we introduce the framework for simulating neutral scalar field collapse in Hayward spacetime.
Section~\ref{section3} presents our key results, describing the dynamical evolution of both matter and geometry as characteristic parameter of scalar field varies from weak to strong regimes. 
In particular, we uncover a scaling relationship between the radius of the inner horizon and the initial profile of the scalar field as the latter approaches a critical threshold.
Finally, we summarize our findings in Section~\ref{section4}.

\section{\label{section2}Methodology}
Let us build the framework for collapse of a massless scalar field in Hayward geometry.
The bulk action is derived from an Einstein-nonlinear electrodynamics model coupled to a massless scalar field, and is given by
\begin{equation}
S = \int {\sqrt { - g} } {d^4}x\left( {\frac{R}{4} + {L^{(1)}} + {L^{(2)}}} \right),
\end{equation}
where
\begin{equation}
{L^{(1)}} =  - \frac{3}{{2s}}\frac{{{{\left( {2{q^2}{\cal F}} \right)}^{3/2}}}}{{{{\left[ {1 + {{\left( {2{q^2}{\cal F}} \right)}^{3/4}}} \right]}^2}}},\quad {L^{(2)}} =  - \frac{1}{2}{g^{\alpha \beta }}{\varphi _{,\alpha }}{\varphi _{,\beta }}.
\end{equation}
Here, $R$ is the Ricci scalar.
${L^{(1)}}$ is the Lagrangian density, which depends on ${\cal F} = {F_{ab}}{F^{ab}}/4$, where ${F_{ab}} = {\partial _a}{{ A}_b} - {\partial _b}{{ A}_a}$ is the electromagnetic-field tensor and ${{ A}_a}$ is the electromagnetic four-potential.
By applying the variational principle to the action with respect to the metric, gauge field, and scalar field, we obtain the following equations of motion:
\begin{equation}
{R_{ab}} - \frac{1}{2}{g_{ab}}R = 2\left( {T_{ab}^{(1)} + T_{ab}^{(2)}} \right),
\end{equation}
\begin{equation}
{\nabla _a}\left( {\frac{{\partial {L^{(1)}}}}{{\partial {\cal F}}}{F^{ab}}} \right) = 0,
\end{equation}
\begin{equation}
\square \varphi   = 0,
\end{equation}
where the corresponding energy-momentum tensors are given by
\begin{equation}\label{Nb11aqwe}
T_{ab}^{(1)} =  - \frac{{\partial {L^{(1)}}}}{{\partial {\cal F}}}{F_{ac}}{F_b}^c + {g_{ab}}{L^{(1)}},\quad T_{ab}^{(2)} = {\varphi _{,a}}{\varphi _{,b}} + {g_{ab}}{L^{(2)}}.
\end{equation}

We consider a general spherically symmetric spacetime described by the ansatz
\begin{equation}\label{Nb11a}
\begin{split}
ds^2 &= 4\, e^{-2\sigma(u,v)} \, du\, dv + r^2(u,v)\, d\Omega^2 \\
     &= e^{-2\sigma(t,x)} \left(-dt^2 + dx^2\right) + r^2(t,x)\, d\Omega^2,
\end{split}
\end{equation}
with \(u\) and \(v\) denoting the null coordinates, defined by $u = (t - x)/2$ and $v = (t + x)/2$.

We adopt the following ansatz for the electromagnetic field
\begin{equation}
{{ A} } = q\cos \theta d\varphi ,
\end{equation}
where $q$ is the magnetic charge.
Consequently, the only non-vanishing component of the electromagnetic field tensor is ${F_{\theta \varphi }} = - q\sin \theta$.

The metric of a static, spherically symmetric Hayward black hole is given by
\begin{equation}\label{Nb11aa}
d{s^2} =  - f(r)d{t^2} + f{(r)^{ - 1}}d{r^2} + {r^2}d{\Omega ^2},
\end{equation}
where
\begin{equation}
f(r) = 1 - \frac{{{q^3}{r^2}}}{{({r^3} + {q^3})s}} = \frac{{(r - {r_ + })(r - {r_ - })(r - {r_n})}}{{{r^3} + {q^3}}},
\end{equation} 
\begin{equation}
s = \frac{{{q^3}r_ + ^2}}{{{q^3} + r_ + ^3}},\quad {r_ - } = \frac{{{q^3} + \sqrt {{q^6} + 4{q^3}r_ + ^3} }}{{2r_ + ^2}},\quad {r_n} = \frac{{{q^3} - \sqrt {{q^6} + 4{q^3}r_ + ^3} }}{{2r_ + ^2}}.
\end{equation}
Here, ${{r_ + }}$ and ${{r_ - }}$ denote the event and inner horizons, respectively.
To examine the nonlinear dynamics of the Hayward geometry under scalar field perturbations, we can write its external metric using Kruskal-like coordinates as follows
\begin{equation}\label{N11asd}
d{s^2} = \frac{{4\rho _0^2{r_n}{r_ + }{r_ - }}}{{({r^3} + {q^3})}}{\left( {\frac{{r - {r_ - }}}{{{r_ - }}}} \right)^{1 - \frac{{{\rho _1}}}{{{\rho _0}}}}}{\left( {\frac{{{r_n} - r}}{{{r_n}}}} \right)^{1 - \frac{{{\rho _2}}}{{{\rho _0}}}}}{e^{ - \frac{r}{{{\rho _0}}}}}( - d{t^2} + d{x^2}) + {r^2}d{\Omega ^2},
\end{equation}
where 
\begin{equation}
{\rho _0} = \frac{{r_ + ^3 + {q^3}}}{{\left( {{r_ + } - {r_ - }} \right)\left( {{r_ + } - {r_n}} \right)}},\quad {\rho _1} = \frac{{r_ - ^3 + {q^3}}}{{\left( {{r_ - } - {r_ + }} \right)\left( {{r_ - } - {r_n}} \right)}},\quad {\rho _2} = \frac{{r_n^3 + {q^3}}}{{\left( {{r_n} - {r_ + }} \right)\left( {{r_n} - {r_ - }} \right)}}.
\end{equation}
$r$ as a function of $(t,x)$ is defined by
\begin{equation}\label{N2}
{t^2} - {x^2} = {e^{\frac{r}{{{\rho _0}}}}}\frac{{{r_ + } - r}}{{{r_ + }}}{\left( {\frac{{r - {r_ - }}}{{{r_ - }}}} \right)^{1 - \frac{{{\rho _1}}}{{{\rho _0}}}}}{\left( {\frac{{{r_n} - r}}{{{r_n}}}} \right)^{1 - \frac{{{\rho _2}}}{{{\rho _0}}}}}.
\end{equation}

The equations of motion for $r$, $\sigma$, and $\varphi$ are
\begin{equation}\label{N11}
r\left( { - {r_{,tt}} + {r_{,xx}}} \right) - r_{,t}^2 + r_{,x}^2 = {{\rm{e}}^{ - 2\sigma }}\left[ {1 - \frac{{3{q^6}{r^2}}}{{s{{\left( {{q^3} + {r^3}} \right)}^2}}}} \right],
\end{equation}
\begin{equation}\label{N11dfg}
 - {\sigma _{,tt}} + {\sigma _{,xx}} + \frac{{{r_{,tt}} - {r_{,xx}}}}{r} + \varphi _{,t}^2 - \varphi _{,x}^2 = {{\rm{e}}^{ - 2\sigma }}\left[ {\frac{{3{q^6}\left( {{q^3} - 2{r^3}} \right)}}{{s{{\left( {{q^3} + {r^3}} \right)}^3}}}} \right],
\end{equation}
\begin{equation}\label{Nb11}
 - \varphi _{,tt}^{} + \varphi _{,xx}^{} + \frac{2}{r}( - {r_{,t}}\varphi _{,t}^{} + {r_{,x}}\varphi _{,x}^{}) = 0,
\end{equation}
supplemented by the constraint equations:
\begin{equation}\label{N9}
r_{, t x}+r_{, t} \sigma_{, x}+r_{, x} \sigma_{, t}+ r \varphi _{, t} \varphi _{, x}=0,
\end{equation}
\begin{equation}\label{N12}
r_{, t t}+r_{, x x}+2\left(r_{, t} \sigma_{, t}+r_{, x} \sigma_{, x}\right)+ r\left(\varphi _{, t}^2+\varphi _{, x}^2\right)=0.
\end{equation}

We specify the initial profile of the scalar field $\varphi$ as a Gaussian wave packet
\begin{equation}\label{N11sdf}
\varphi (0,x) = C{e^{ - \frac{{{{\left( {x - {x_0}} \right)}^2}}}{D}}}.
\end{equation}
This choice provides a smooth, localized, finite-energy perturbation and allows for a controlled parametrization of the initial data in terms of amplitude $C$, width $D$, and central position $x_0$.
Gaussian wave packets are routinely adopted in numerical studies of gravitational collapse and critical phenomena~\cite{Choptuik:1992jv, Gundlach:1999cu, Goldwirth:1987nu,Winicour:2008vpn,Garfinkle:1994jb}.
A Gaussian profile is used here purely as a convenient representative of a broad class of smooth and localized initial perturbations, without implying any special or preferred physical configuration.
In principle, any smooth and localized initial data with finite energy can be adopted.
As will be discussed below, the subsequent evolution of the system, particularly the onset of mass inflation and the contraction of the inner horizon, is mediated in part through scalar energy transport and its nonlinear backreaction, rather than by the detailed functional form of the initial scalar profile.

Building upon the Hayward metric expressed in Kruskal-like coordinates~({\ref{N11asd}}), we impose a gauge condition on the function $\sigma$ as below
\begin{equation}
{\left. {{e^{ - 2\sigma }}} \right|_{t = 0}} = \left. {{e^{ - 2\sigma }}} \right|_{t = 0}^{{\rm{Hayward}}} = \frac{{4\rho _0^2{r_n}{r_ + }{r_ - }}}{{({r^3} + {q^3})}}{\left( {\frac{{r - {r_ - }}}{{{r_ - }}}} \right)^{1 - \frac{{{\rho _1}}}{{{\rho _0}}}}}{\left( {\frac{{{r_n} - r}}{{{r_n}}}} \right)^{1 - \frac{{{\rho _2}}}{{{\rho _0}}}}}{e^{ - \frac{r}{{{\rho _0}}}}},
\end{equation}
where $r$ is obtained from Eq.~({\ref{N2}}).
Combining Eqs.~(\ref{N11}) and (\ref{N12}), we obtain the initial radial profile for $r$ as below
\begin{equation}\label{N99}
{r_{,xx}} =  - {r_{,t}}{\sigma _{,t}} - {r_{,x}}{\sigma _{,x}} + \frac{{r_{,t}^2 - r_{,x}^2}}{{2r}} - \frac{r}{2}(\varphi _{,x}^2 + \varphi _{,t}^2) + \frac{{{{\rm{e}}^{ - 2\sigma }}}}{{2r}}\left[ {1 - \frac{{3{q^6}{r^2}}}{{s{{\left( {{q^3} + {r^3}} \right)}^2}}}} \right].
\end{equation}
The initial conditions are set to be time-symmetric ${r_{,t}} = {\sigma _{,t}} = {\varphi  _{,t}} = 0$, so that the constraint equation~({\ref{N9}}) is automatically satisfied.
The initial profile of $r$ can then be computed numerically using the fourth-order Runge–Kutta method.
The computational domain for $x$ is set as $\left[ { - {x_a},{x_a}} \right]$.
The boundary values of $r$, $\sigma $, and $\varphi  $ at $x =  \pm {x_a}$ are typically obtained through extrapolation. 
Provided that the boundaries are placed sufficiently far from the central region of interest, they do not significantly affect the dynamics within the domain that we are concerned with. 
For the time evolution, we adopt the leapfrog method, which requires initial data at both $t = 0$ and $t = \Delta t$.
The data at $t = \Delta t$ are obtained via a Taylor expansion using the initial data at $t = 0$.
Subsequent evolution is then carried out by iteratively applying the finite difference scheme.
For further details regarding the numerical implementation, please refer to Ref.~\cite{Guo:2015laa}.

\section{\label{section3}Results}
We simulate collapse of a scalar field with varying amplitudes in Hayward geometry, tracking the evolution of the metric functions and the field itself. 
For sufficiently strong perturbations, we check whether the final state of the black hole remains regular or develops a central spacelike singularity. 
In the near-critical regime, we investigate how the initial profile of scalar field influences the radius of the inner horizon, seeking to characterize their quantitative correlation.

\subsection{\label{sectiona}Collapse of a weak scalar field in Hayward geometry}
In this subsection, the scalar field carries less energy, which is set as~({\ref{N11sdf}}), with $C=0.045$, $D=1$, and ${x_0} = 1$.
In Hayward geometry, we impose $q=0.5$ and ${r_ + } = 3$.
The spatial domain is set as $\left[ { - 5,5} \right]$.

\begin{figure*}[htbp]
\centering
\includegraphics[scale=0.26]{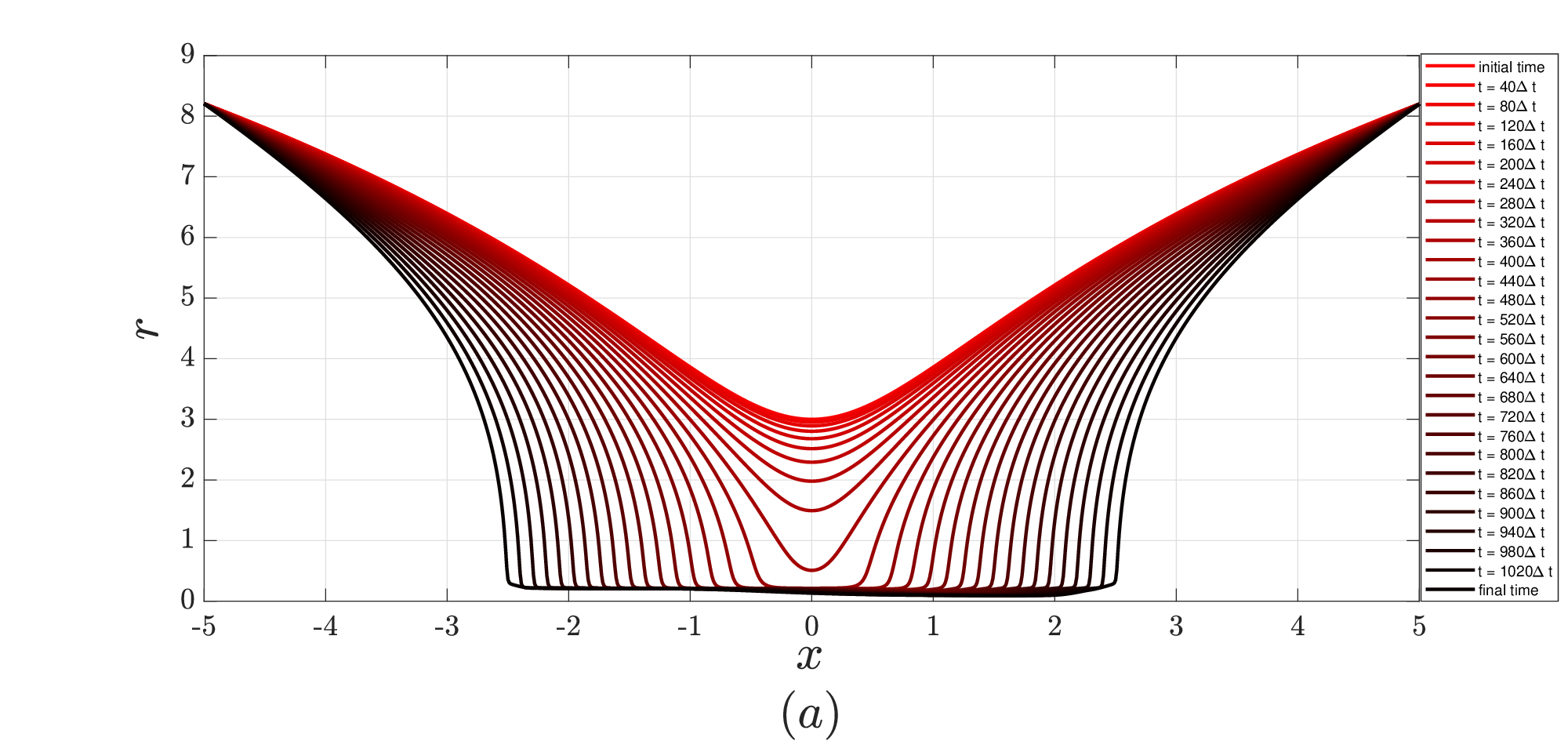}
\includegraphics[scale=0.26]{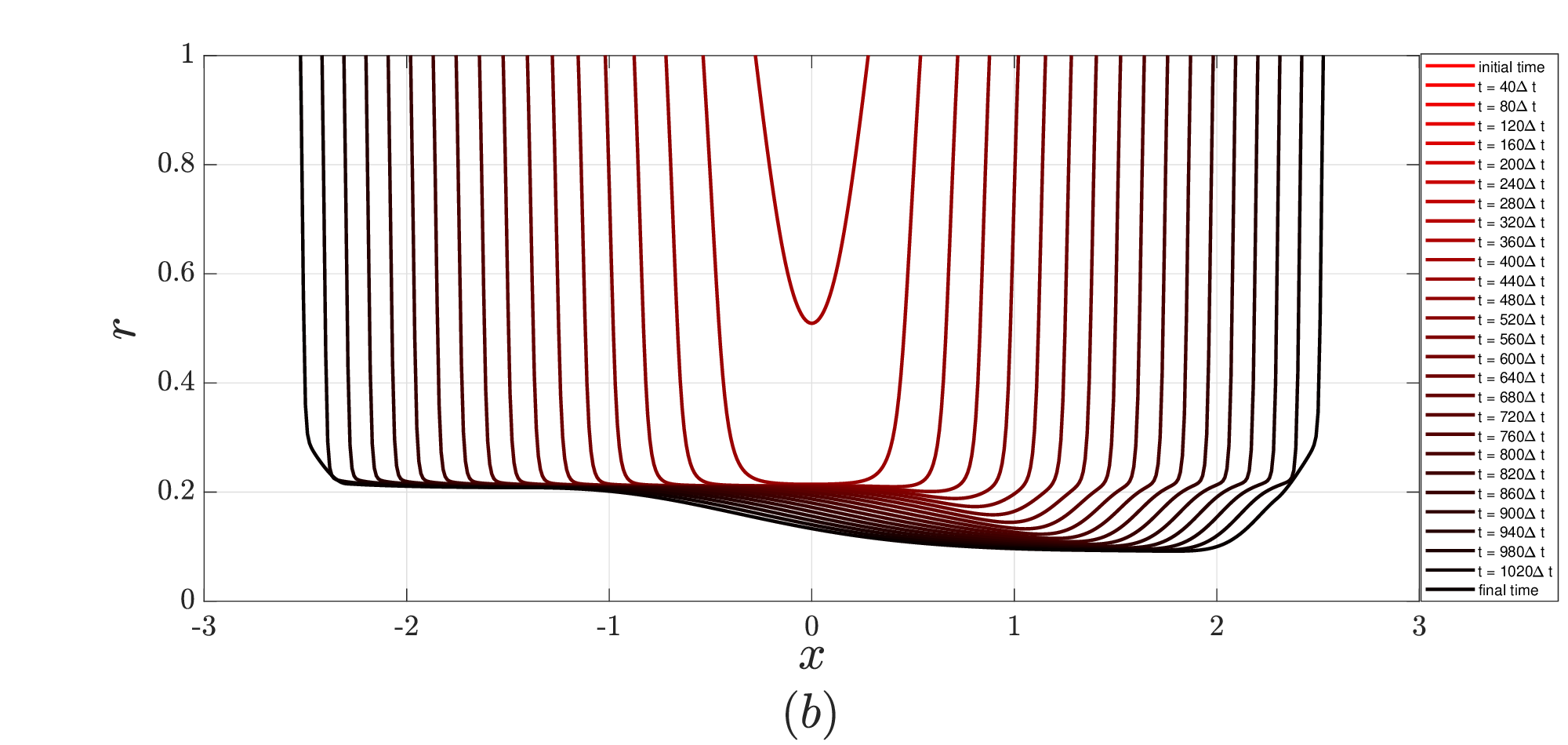}
\includegraphics[scale=0.26]{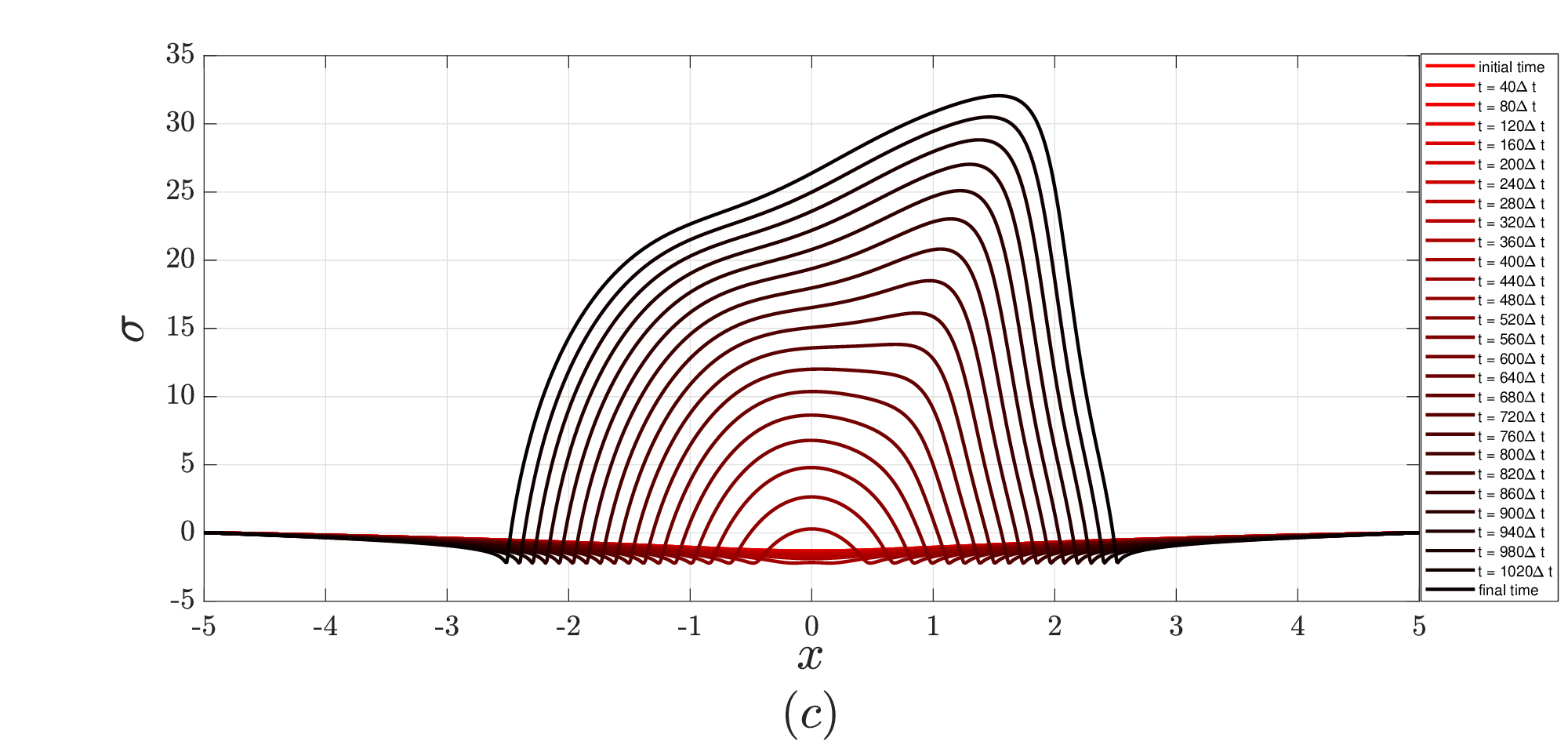}
\includegraphics[scale=0.26]{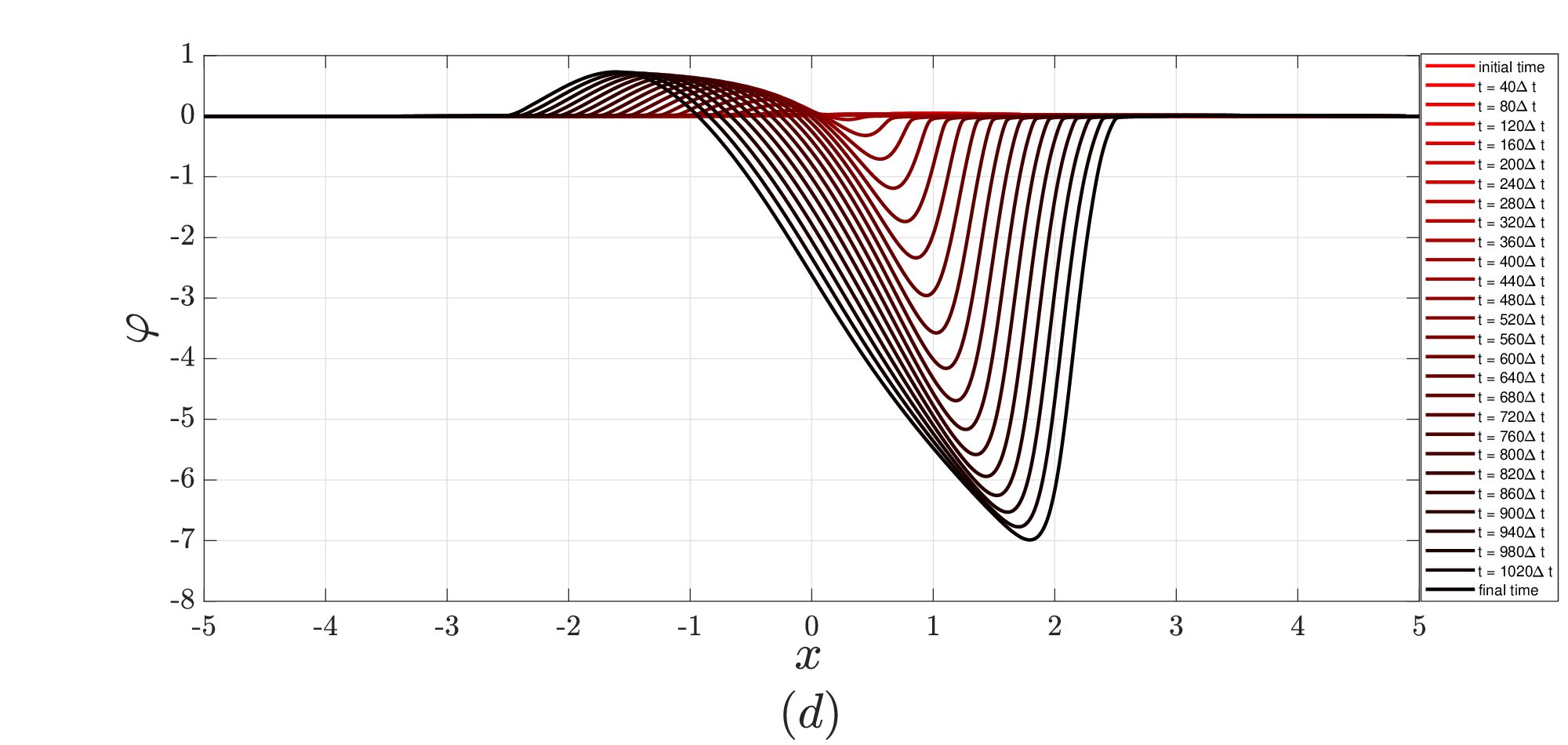}
\caption{ The evolutions for $r$, $\sigma$ and $\varphi $ under a weak scalar field with $C=0.045$.
(b) A detailed view of the evolution of $r$, illustrating the influence of the scalar field on its dynamics.
Each line represents the variation of $r$, $\sigma$, and $\varphi $ with respect to $x$ at fixed time.
}\label{Fig1}
\end{figure*}

\begin{figure*}[htbp]
\centering
\includegraphics[scale=0.29]{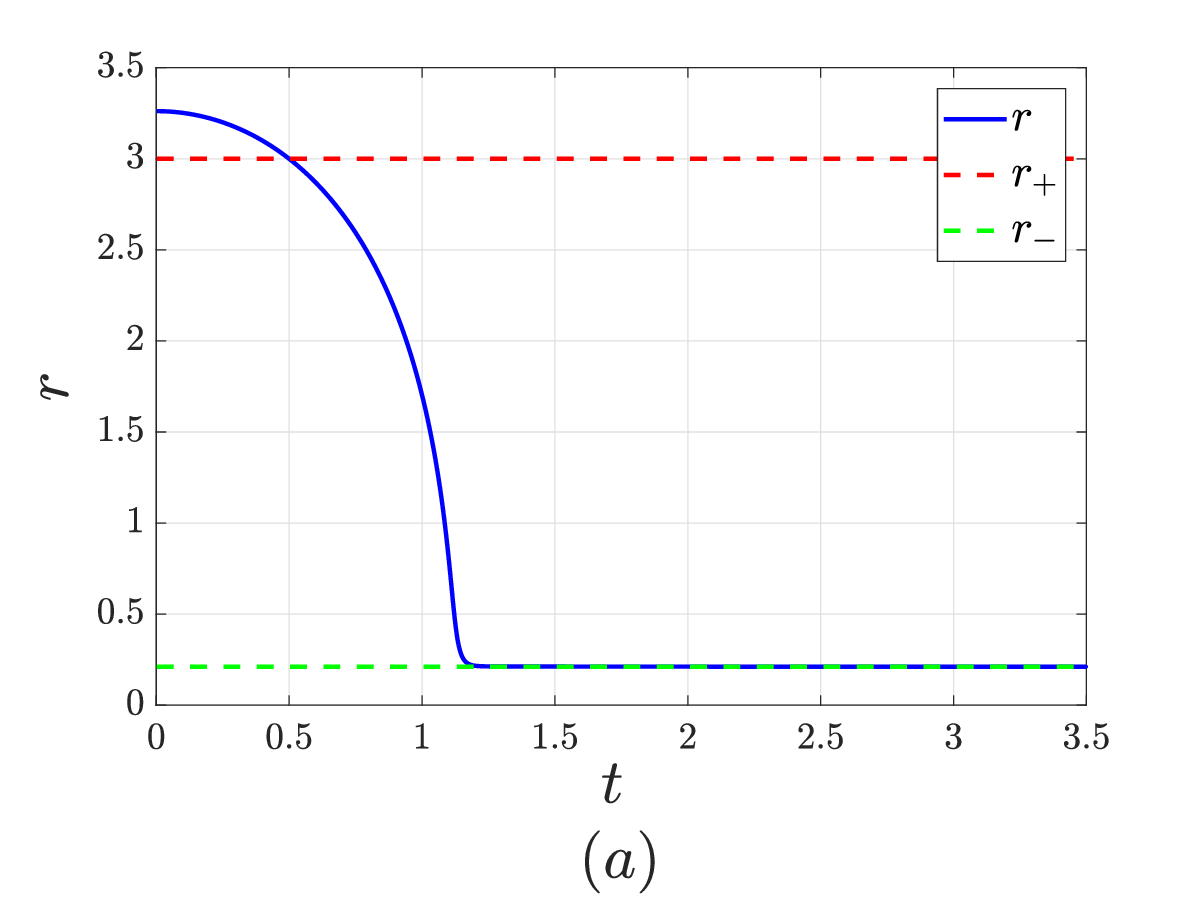}
\includegraphics[scale=0.29]{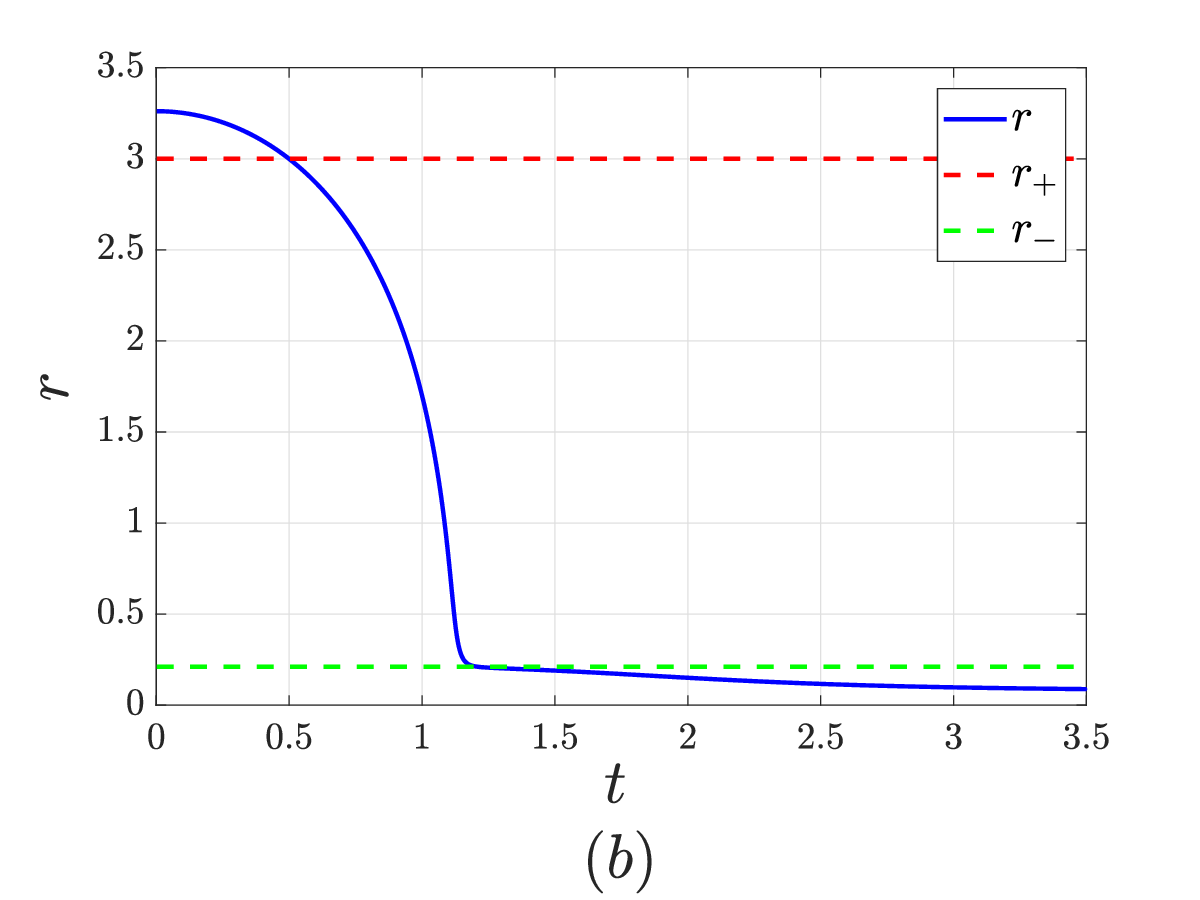}
\includegraphics[scale=0.29]{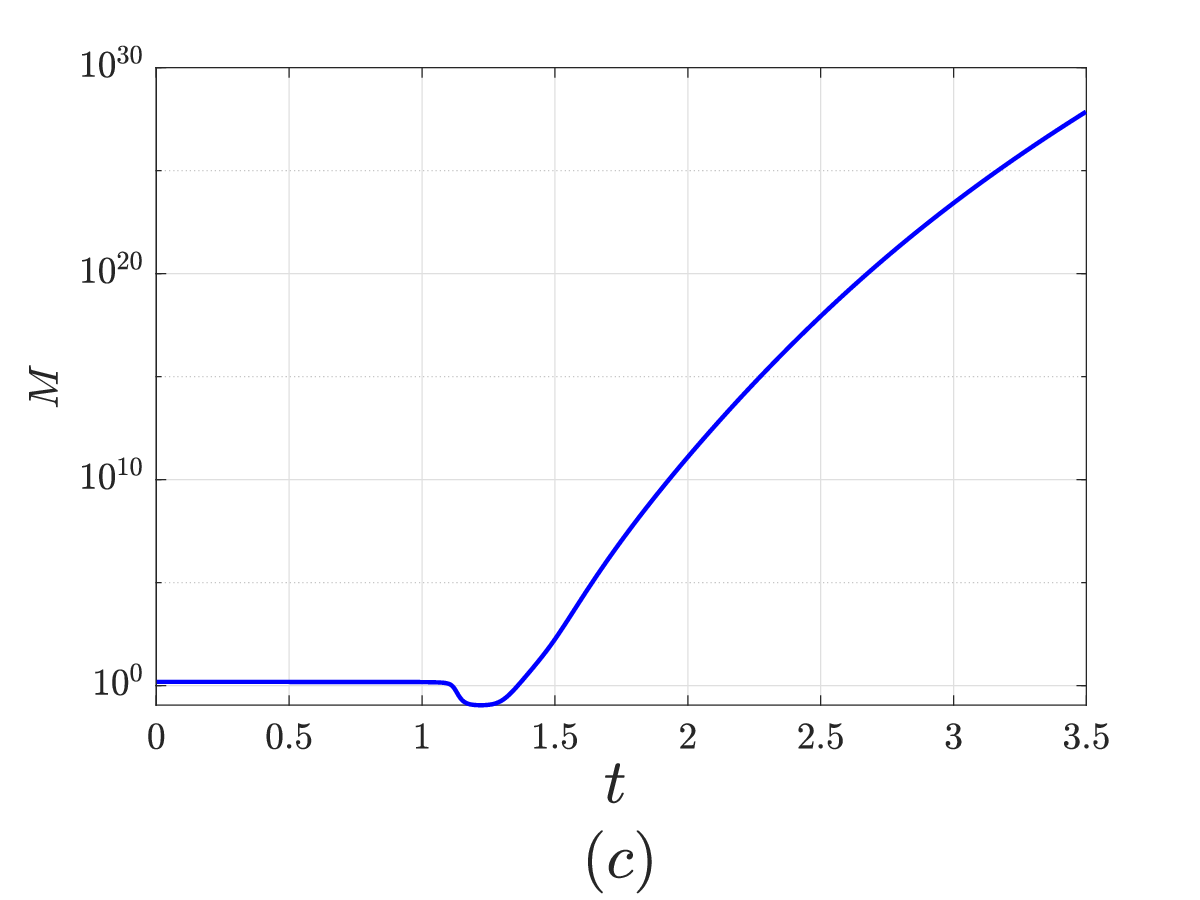}
\caption{ Evolution of black hole characteristics along the $x=0.5$ slice.
(a) Evolution of $r$ with $C = 0$, recovering the original Hayward black hole solution where ${r_ + }$ and ${r_ - }$ denote the radii of the outer and inner horizons of the original Hayward black hole respectively.
(b) Evolution of $r$ with $C  = 0.045$, showing the inner horizon’s contraction due to the scalar field.
(c) Evolution of the Misner–Sharp mass $M$, exhibiting divergence over time for $C  = 0.045$.
}\label{Fig2}
\end{figure*}

The evolutions of the metric components $r$, $\sigma$, and the scalar field $\varphi $ are shown in Fig.~\ref{Fig1}.
The transition from the initial to the final moment in the numerical simulation is represented by a color gradient that ranges from red to black.
As shown in Fig.~\ref{Fig1}(a)-(b), $r$ gradually decreases over time, but the final evolution remains at a non-zero value. 
Moreover, the amplitude of the scalar field influences the inner horizon of the black hole.
As shown in Fig.~\ref{Fig2}(a), in the absence of the scalar field, the spacetime evolution follows the Hayward geometry, with the radial coordinate $r$ approaching the inner horizon of the Hayward black hole.
When the scalar field is added, $r$ evolves to a value smaller than the inner horizon of the Hayward black hole. 
This behavior indicates that an increase in the strength of the scalar field can accelerate the contraction of the inner horizon.
Furthermore, as shown in Figs.~\ref{Fig1}(c)-(d), the temporal evolution manifests progressive intensification in both the variables $\sigma$ and $\varphi $.
Nevertheless, both quantities remain bounded, showing no tendency toward divergence. 
These observations suggest that the internal structure of the black hole undergoes subtle modifications, potentially impacting its mass, energy distribution, and other related physical properties.
In particular, the Misner–Sharp mass function can be defined as follows:
\begin{equation}\label{N00}
{g^{\mu \nu }}{r_{,\mu }}{r_{,\nu }} = {e^{2\sigma }}( - r_{,t}^2 + r_{,x}^2) \equiv 1 - \frac{{2M(t,x)}}{r}.
\end{equation}
As shown in Fig.~\ref{Fig2}(c), the divergence of the Misner–Sharp mass near the inner horizon provides a characteristic signature of mass inflation.

\subsection{Collapse of a strong scalar field in Hayward geometry}
\begin{figure*}[htbp]
\centering
\includegraphics[scale=0.26]{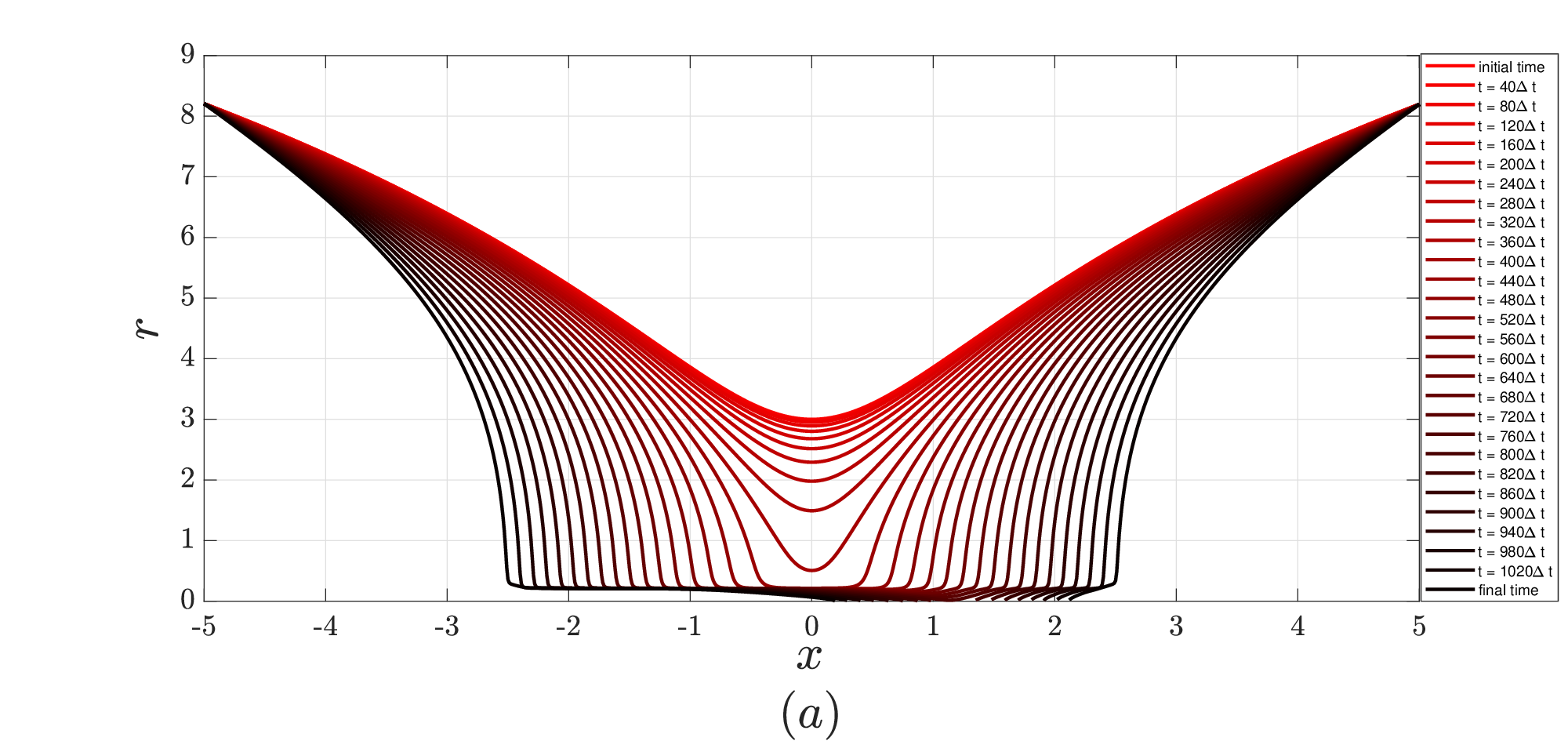}
\includegraphics[scale=0.26]{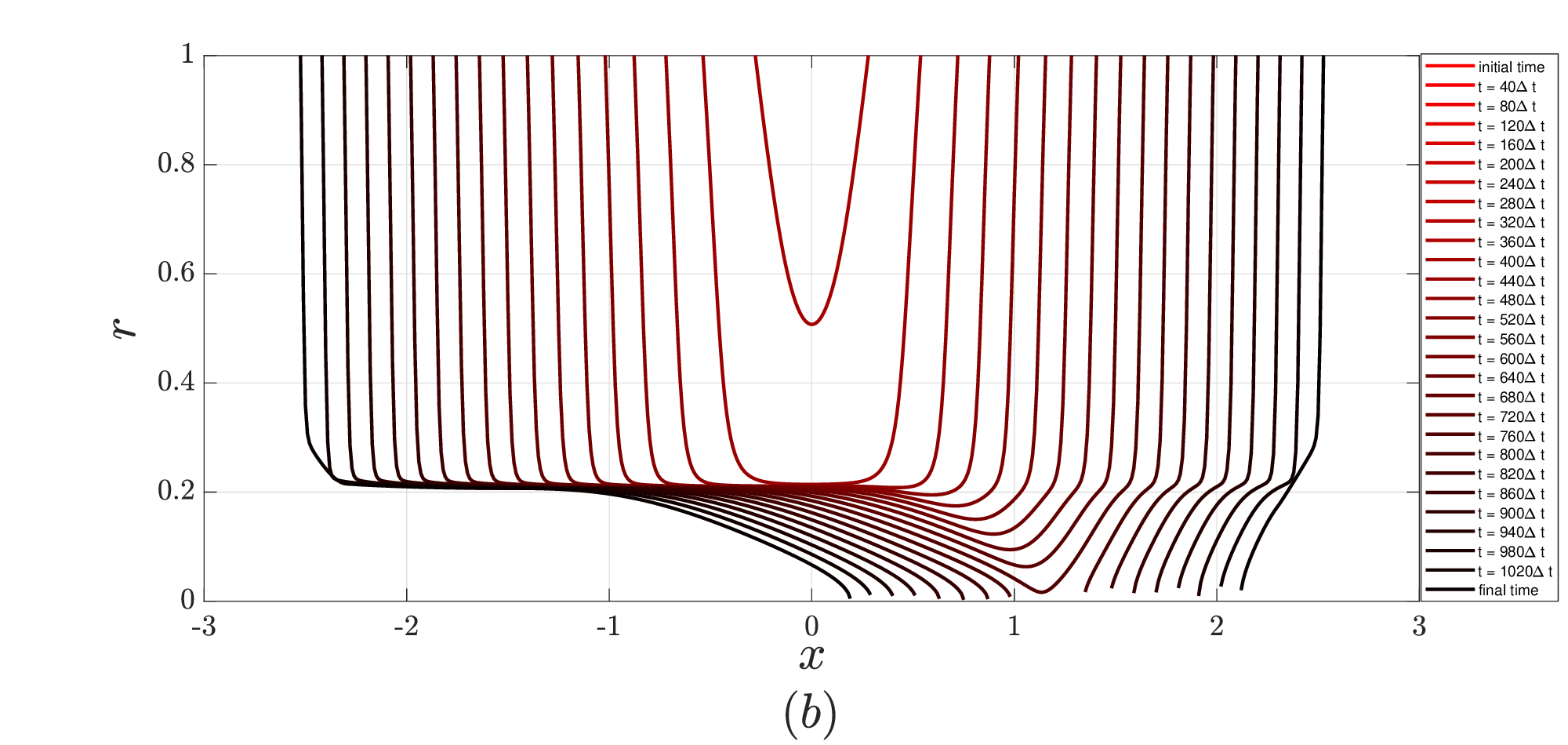}
\includegraphics[scale=0.26]{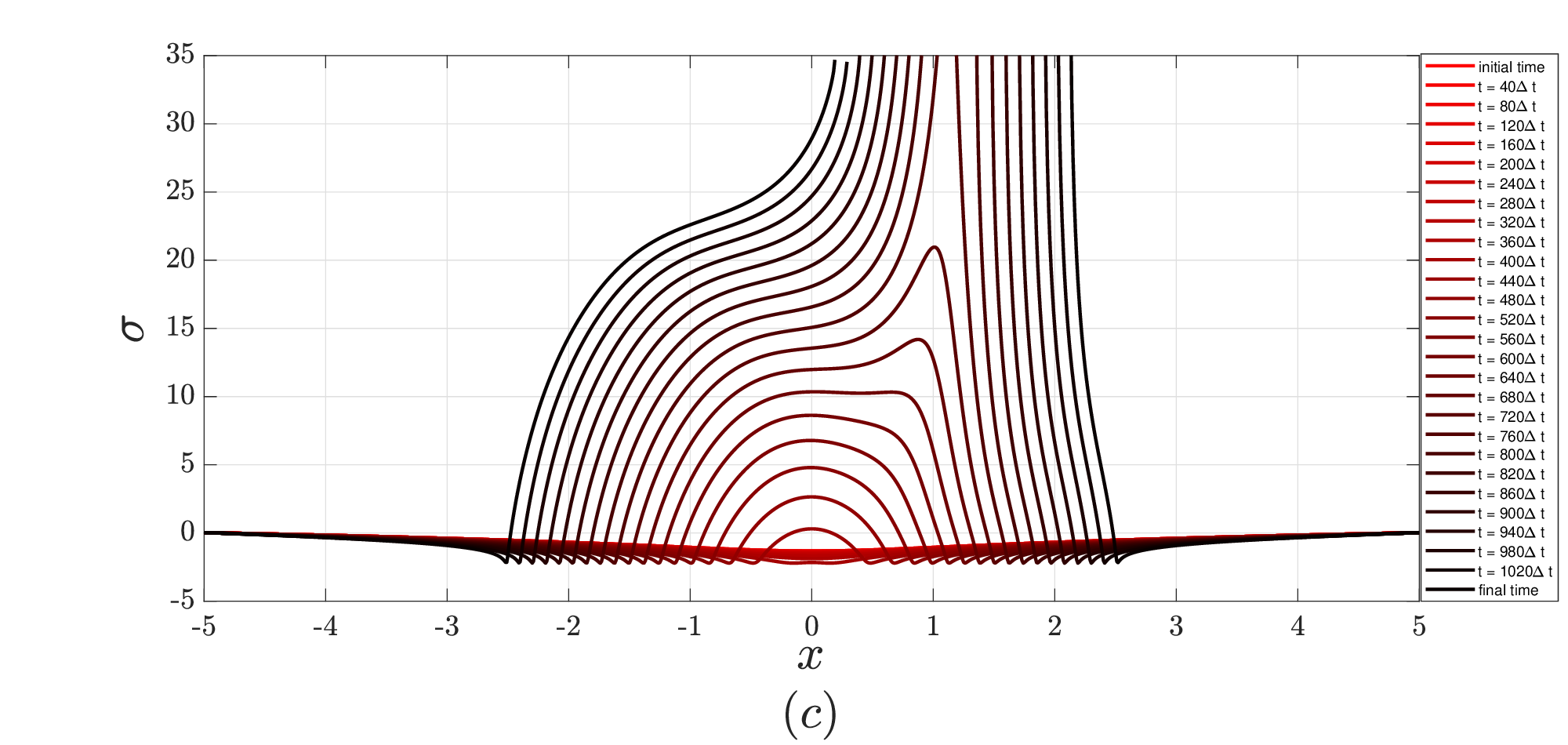}
\includegraphics[scale=0.26]{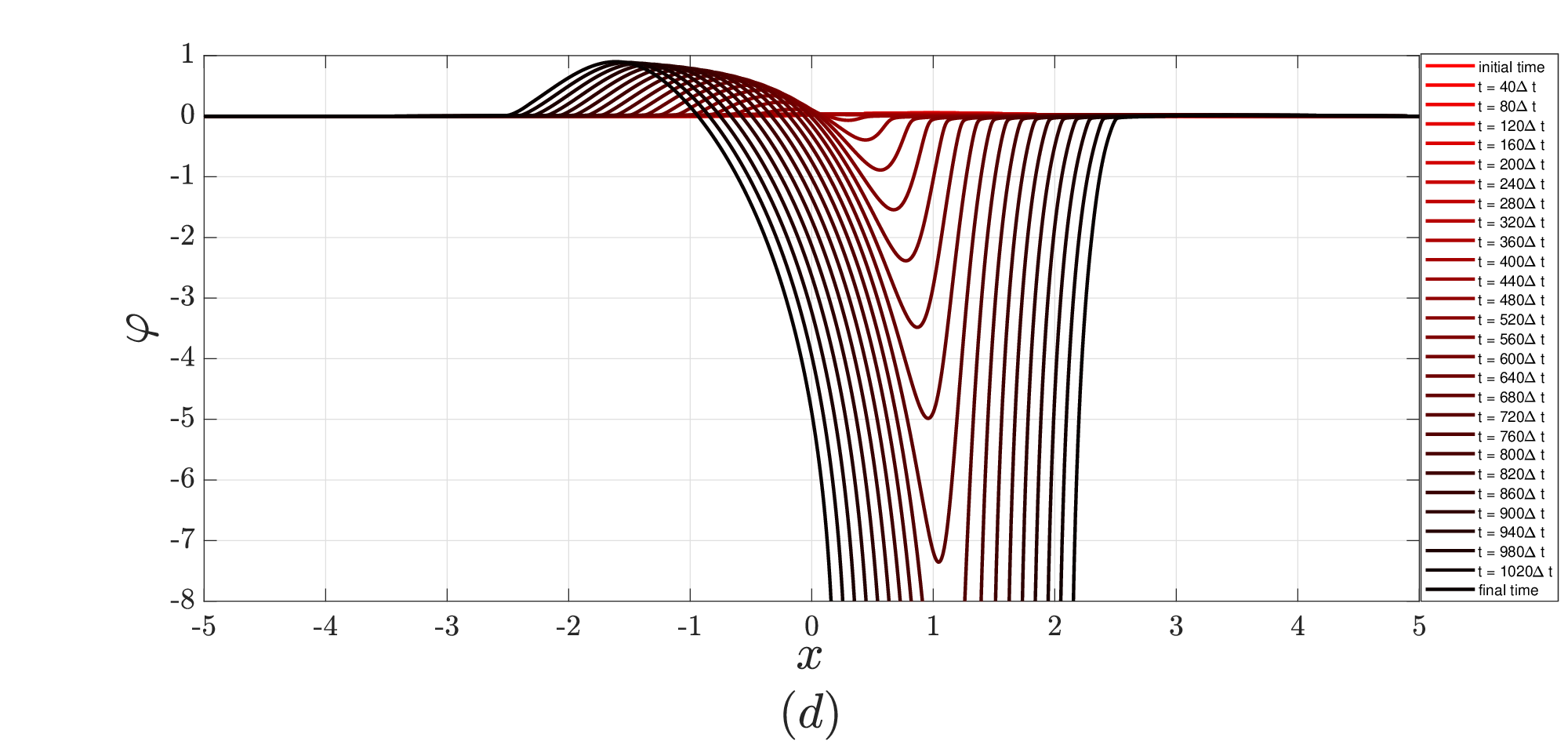}
\caption{ Evolutions for $r$, $\sigma$ and $\varphi $ in a strong scalar field with $C=0.055$.
(b) A detailed view of the evolution of $r$, illustrating the influence of the scalar field on its dynamics.
Each line represents the variation of $r$, $\sigma$, and $\varphi $ with respect to $x$ at fixed time.
}\label{Fig3}
\end{figure*}

\begin{figure}[!]
\centering
\includegraphics[scale=0.35]{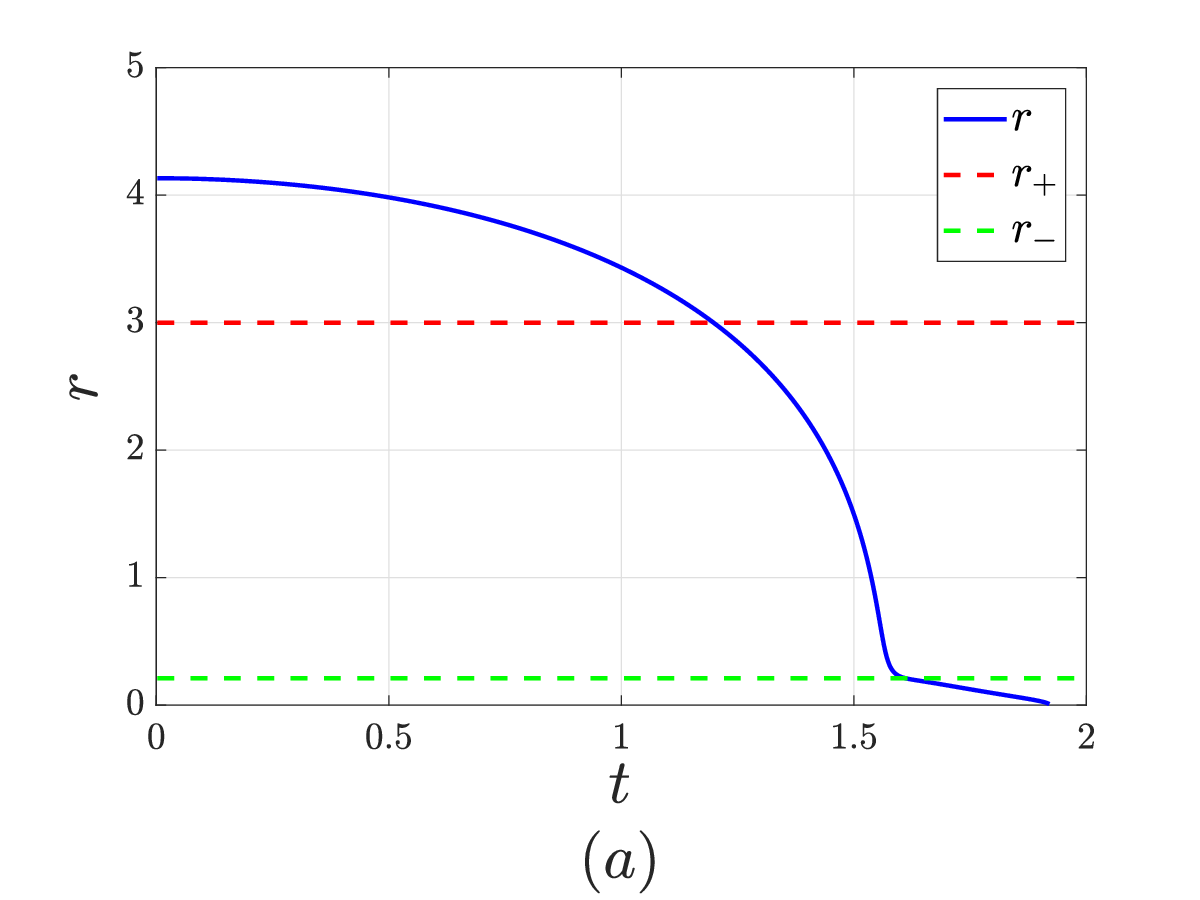}
\includegraphics[scale=0.35]{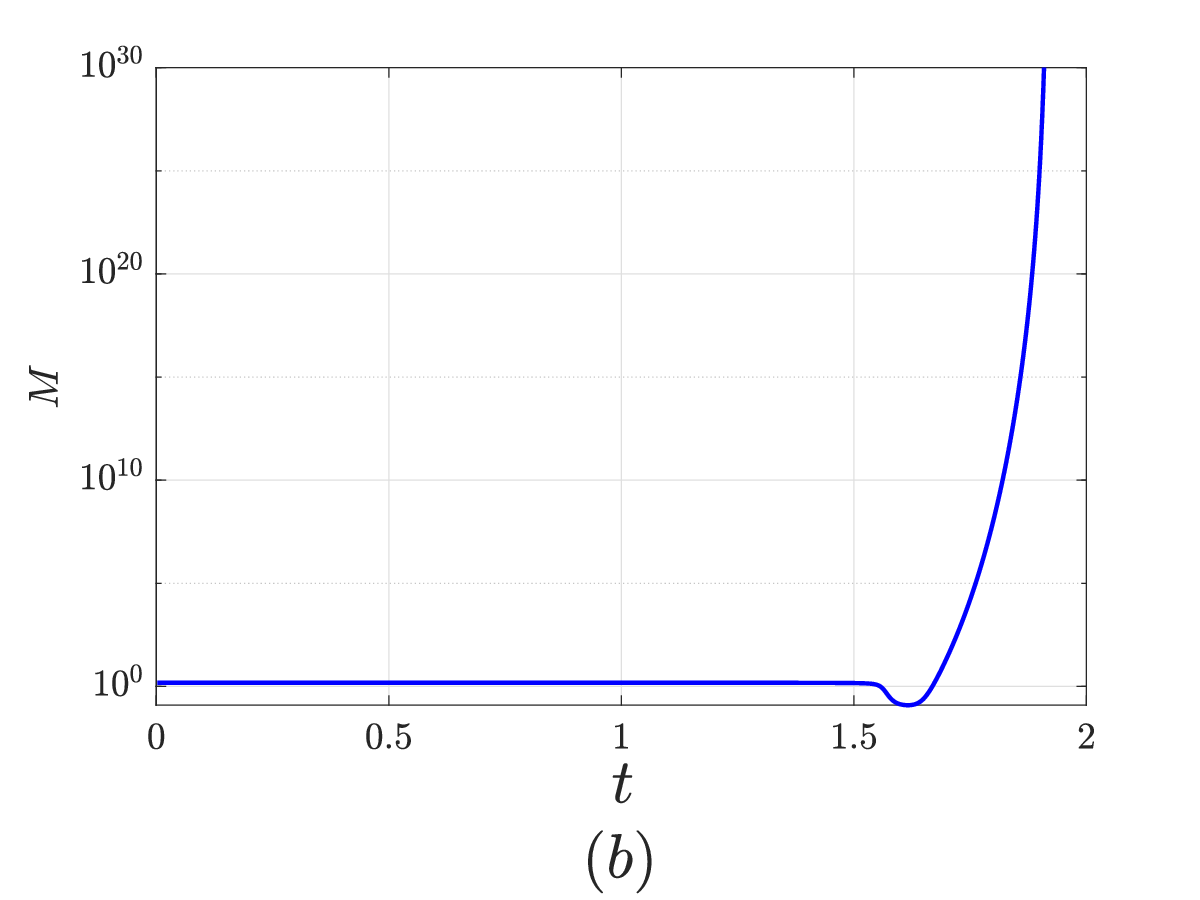}
\includegraphics[scale=0.35]{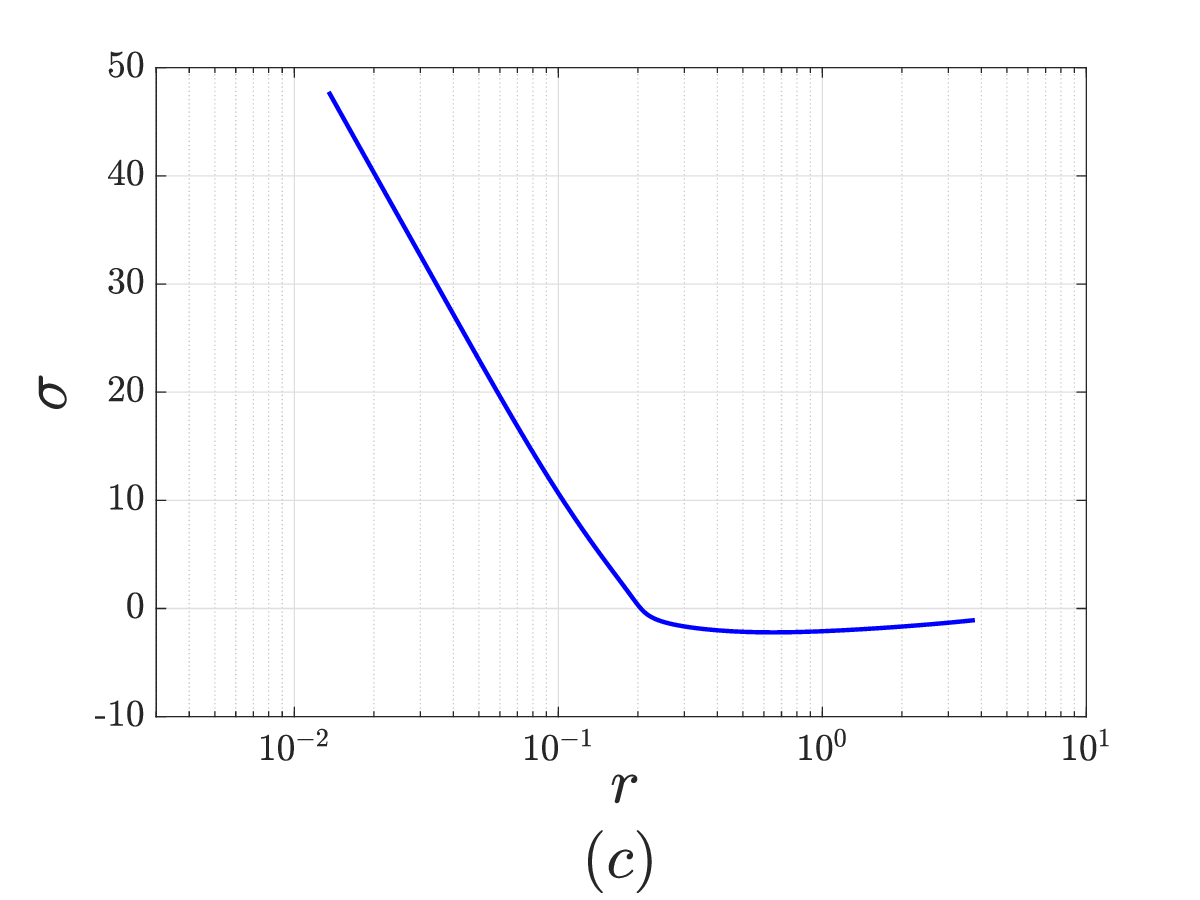}
\includegraphics[scale=0.35]{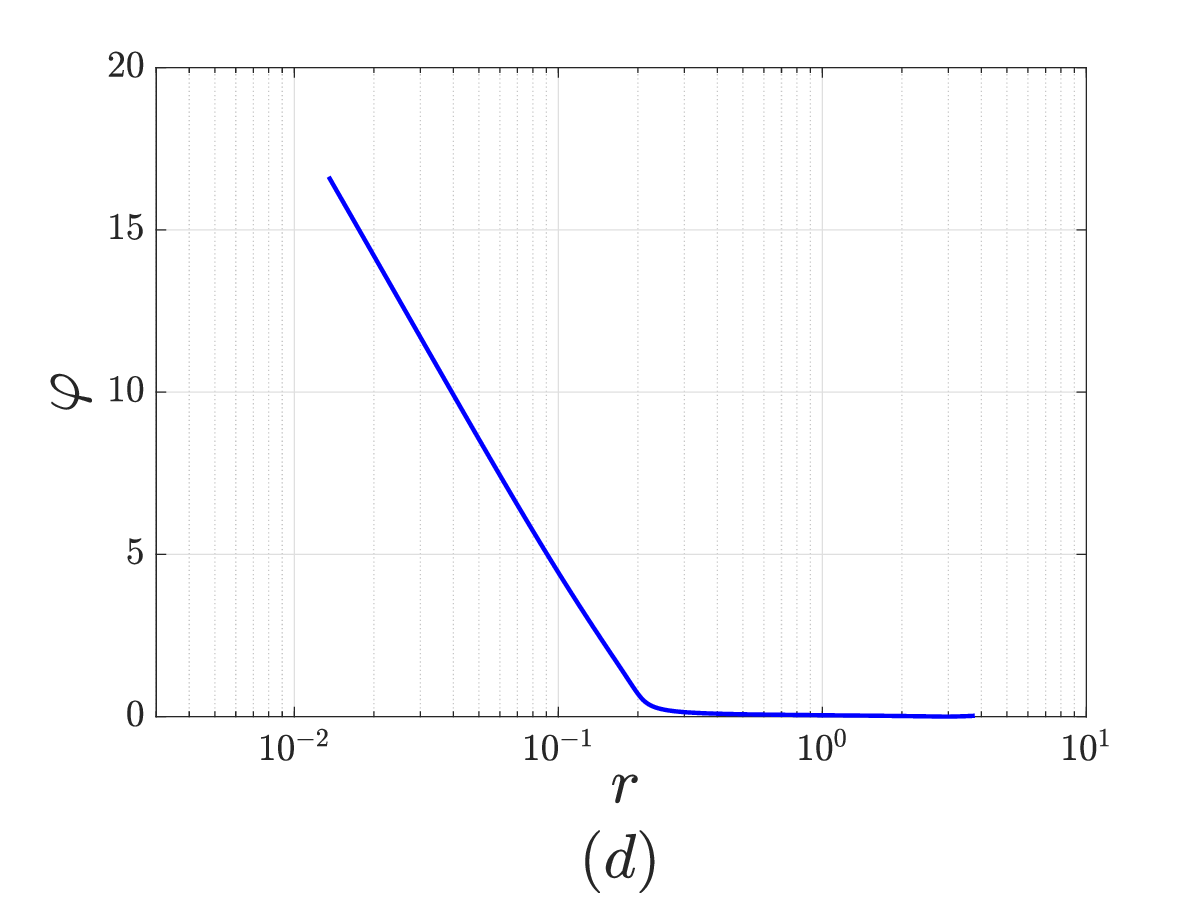}
\caption{ Evolution of black hole characteristics along the $x=1.2$ slice.
(a) Evolution of $r$ with $C  = 0.055$, illustrating the disappearance of the inner horizon. 
Here, ${r_ + }$ and ${r_ - }$ denote, respectively, the radii of the outer and inner horizons of the original Hayward black hole.
(b) Evolution of the Misner–Sharp mass $M$, highlighting the mass inflation phenomenon.
(c) and (d): Evolution of $\sigma$ and $\varphi $, both of which diverge as $r$ approaches zero.
}\label{Fig4}
\end{figure}

As a comparison, we simulate collapse of a strong scalar field in Hayward geometry in this subsection.
As illustrated in Fig.~\ref{Fig3}, $r$ decreases monotonically over time, ultimately shrinking to zero in regions where the scalar field becomes sufficiently strong.
Both the metric function $\sigma$ and the scalar field $\varphi$ exhibit increasing trends during evolution.
They, however, diverge once the scalar field surpasses a critical strength, indicating the onset of dynamical instability. 
Such instability may trigger extreme physical phenomena, fundamentally altering the spacetime structure.
To probe the dynamics more closely, we focus on a spatial slice at $x=1.2$ (see Fig.~\ref{Fig4}). 
As anticipated, the inner horizon contracts to zero rapidly. 
Concurrently, the divergences of $\sigma$ and $\varphi$ induce the mass inflation as $r$ approaches zero. 
These abrupt, nonlinear changes suggest that the core dynamics of the black hole could lead to the formation of a new singularity.

\begin{figure*}[!]
\centering
\includegraphics[scale=0.4]{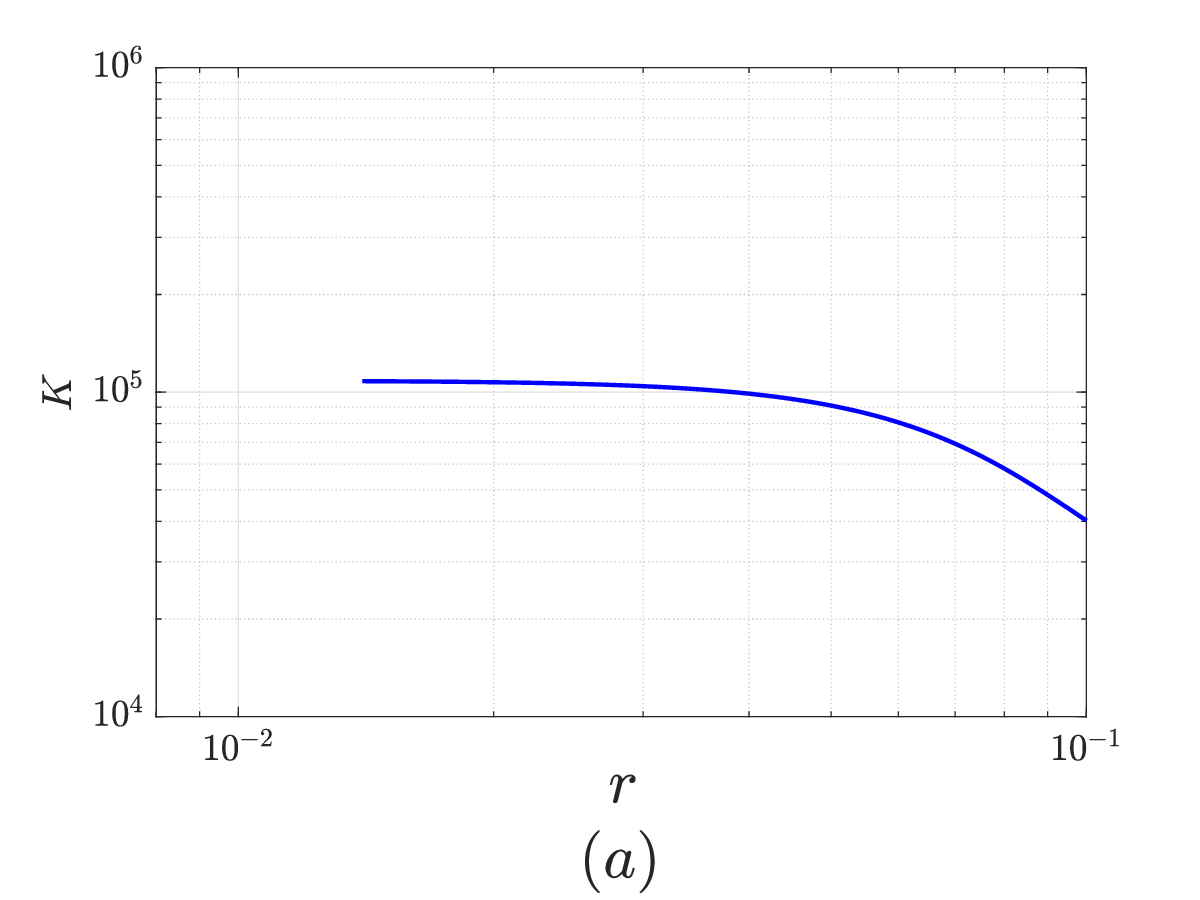}
\includegraphics[scale=0.4]{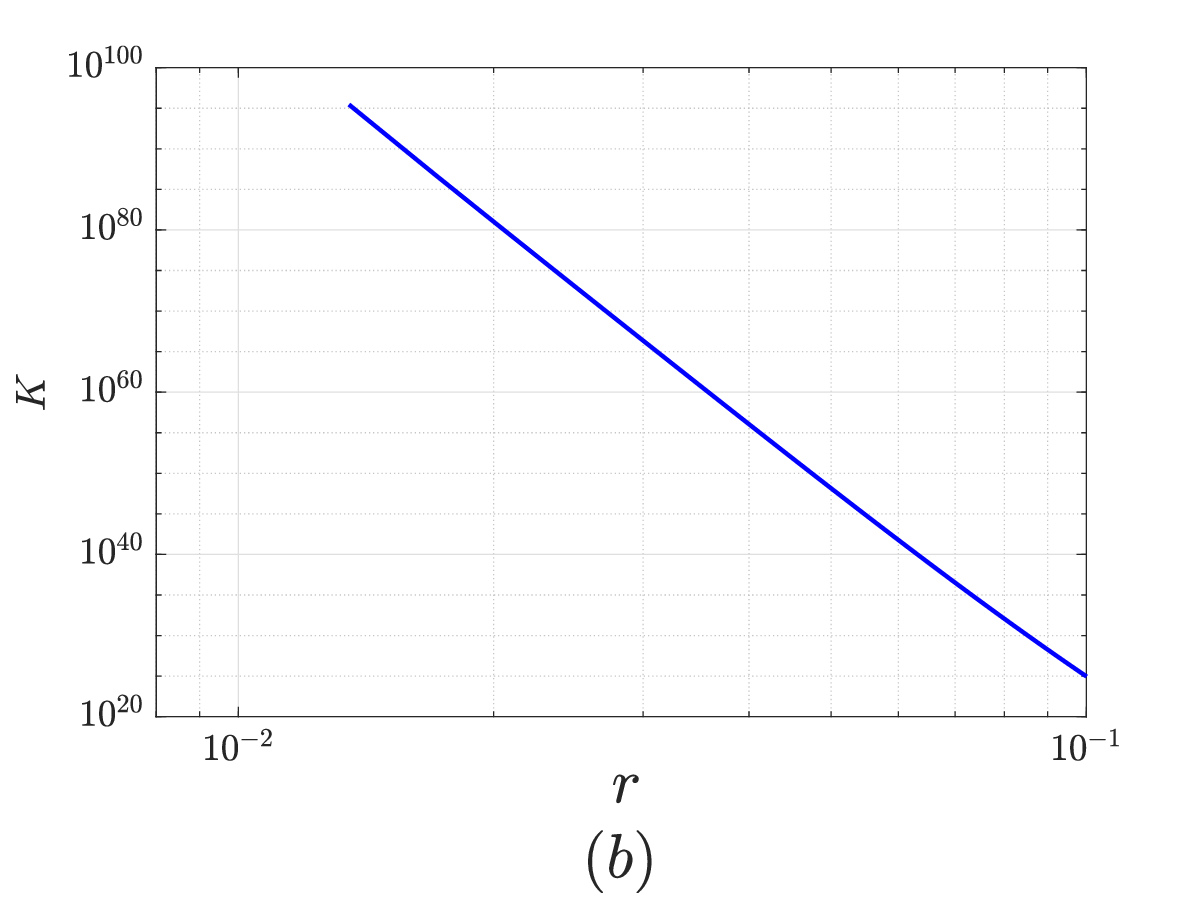}
\caption{Kretschmann scalar $K$ as $r$ approaches zero. (a) Initial state. (b) Final state.
}\label{Fig5}
\end{figure*}

A thorough analysis of the curvature invariants is performed to verify the predicted emergence of novel singularities.
The Kretschmann scalar for the Hayward geometry takes the explicit form:
\begin{equation}\label{N00}
\begin{split}
K = \frac{4e^{4\sigma}}{r^4} \bigg[ & e^{-4\sigma} + r^4 (\sigma_{,tt} - \sigma_{,xx})^2 + 2e^{-2\sigma} (r_{,t}^2 - r_{,x}^2) + (r_{,t}^2 - r_{,x}^2)^2 \\
& + 2r^2 (r_{,xx}^2 + r_{,tt}^2 - 2r_{,xt}^2) + 4r^2 (r_{,t}^2 \sigma_{,t}^2 + r_{,x}^2 \sigma_{,x}^2 - r_{,t}^2 \sigma_{,x}^2 - r_{,x}^2 \sigma_{,t}^2) \\
& + 4r^2 ( r_{,xx} r_{,x} \sigma_{,x} + r_{,tt} r_{,t} \sigma_{,t} + r_{,tt} r_{,x} \sigma_{,x} + r_{,xx} r_{,t} \sigma_{,t} - 2r_{,xt} r_{,t} \sigma_{,x} - 2r_{,xt} r_{,x} \sigma_{,t} ) \bigg].
\end{split}
\end{equation}
Figure~\ref{Fig5} illustrates the behavior of the Kretschmann scalar $K$ as $r$ approaches zero in both the original Hayward geometry and the final state. 
In the initial case, $K$ remains finite at the origin, reflecting its nonsingular central core. 
In contrast, $K$ diverges at the origin in our model, signaling the formation of a genuine curvature singularity.

\begin{figure*}[!]
\centering
\includegraphics[scale=0.29]{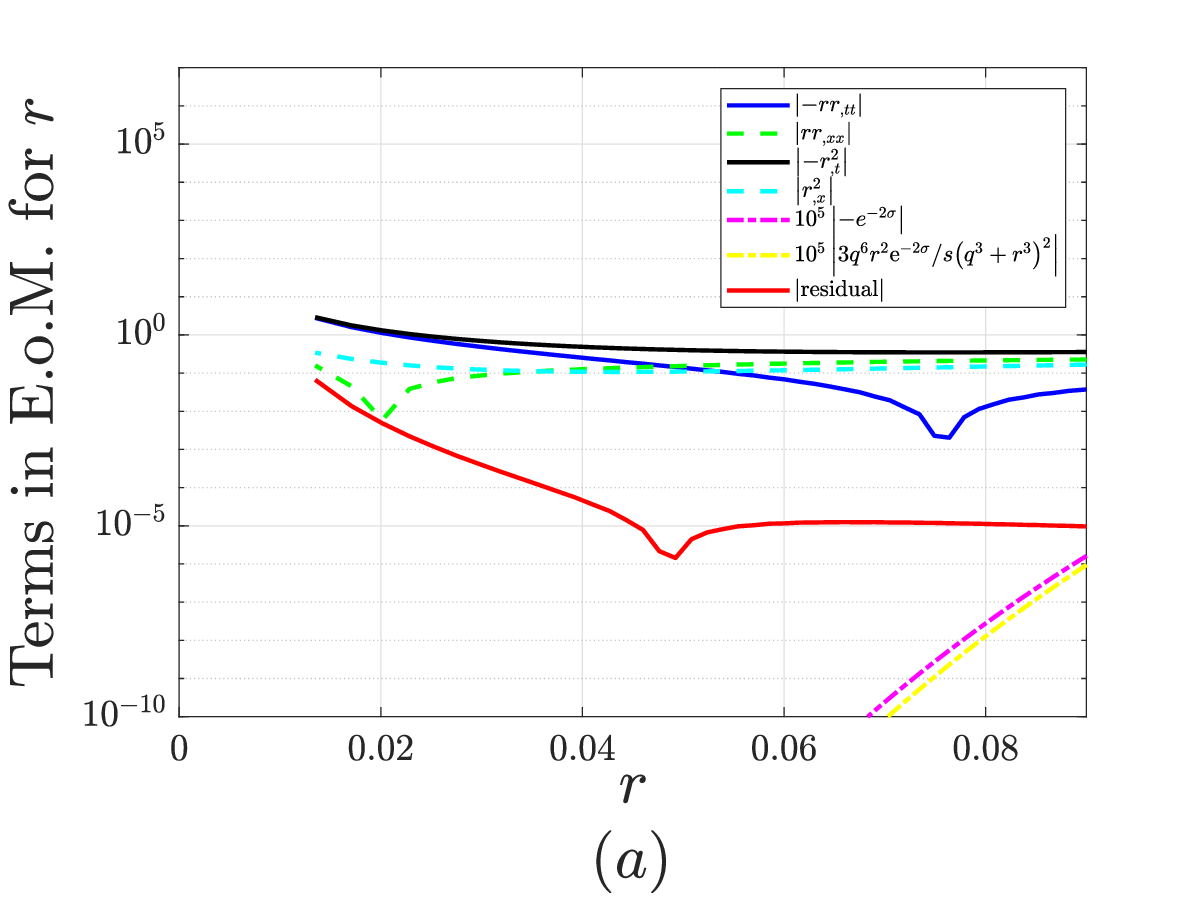}
\includegraphics[scale=0.29]{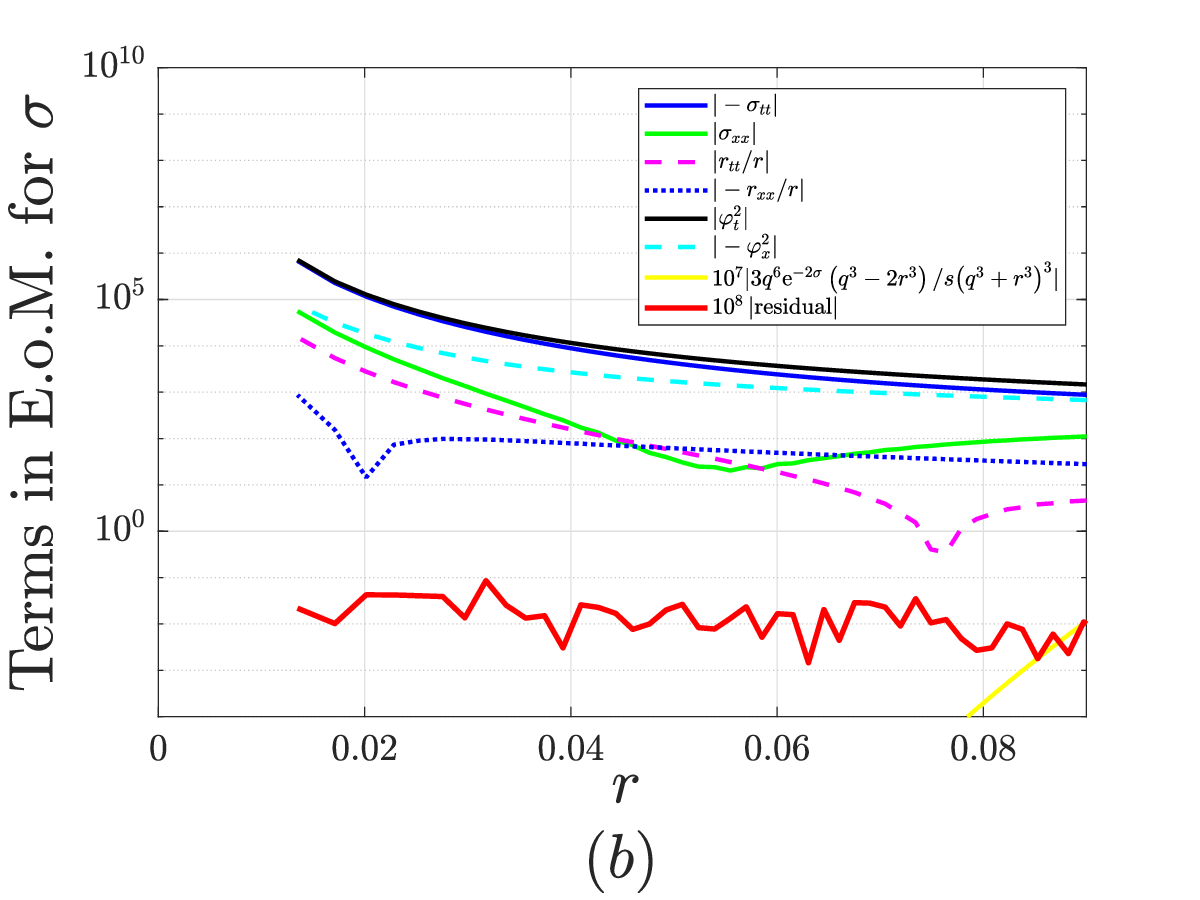}
\includegraphics[scale=0.29]{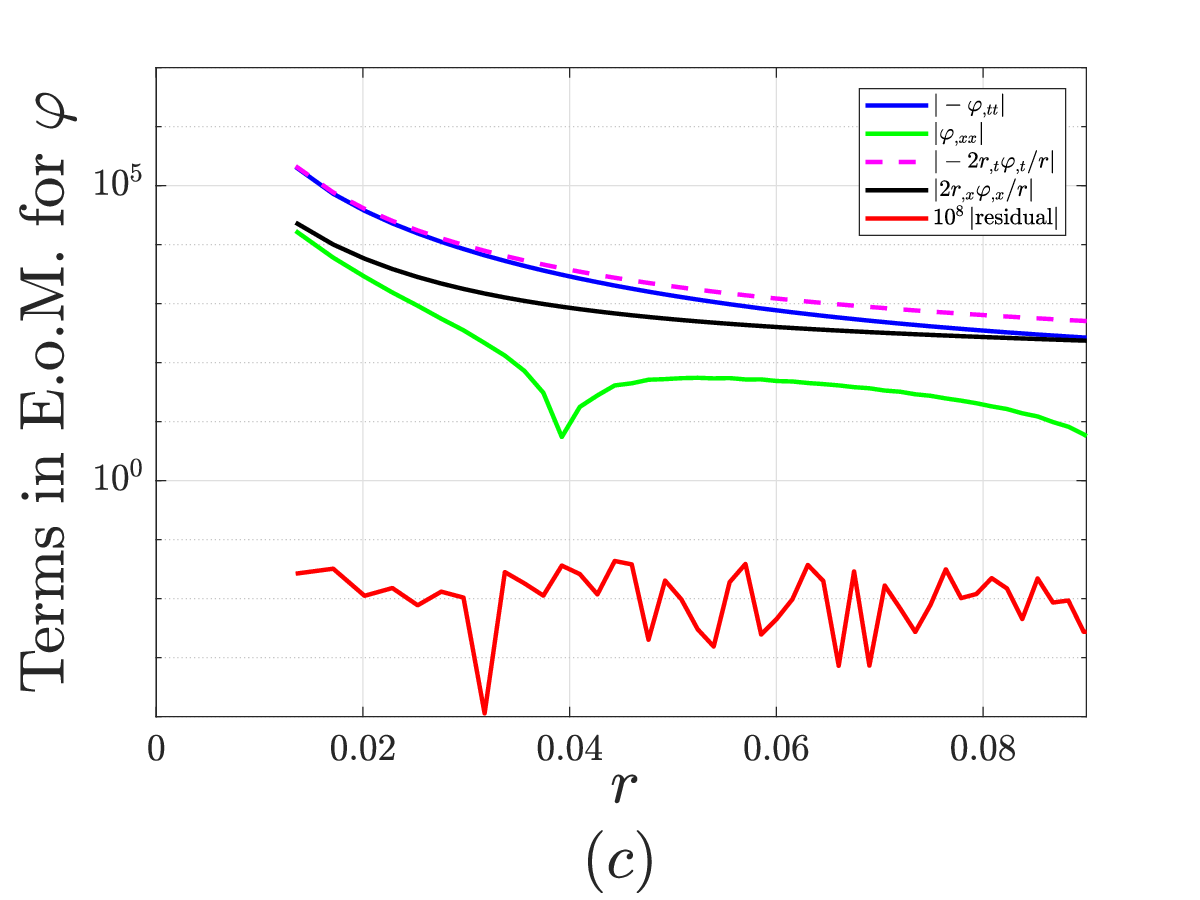}
\caption{Contributions of individual terms in the dynamical equations for $r$ (\ref{N11}), $\sigma$ (\ref{N11dfg}) and $\varphi $ (\ref{Nb11}) as $r$ approaches zero along the slice of $x = 1.2$. 
}\label{Fig6}
\end{figure*}

To probe the spacetime dynamics near the singularity, we systematically track the asymptotic behavior of all terms in the equations of motion for $r$, $\sigma$, and $\varphi $ as $r$ approaches zero in Fig.~\ref{Fig6}.
It is observed that the contributions from the terms $| {3{q^6}{r^2}{{\rm{e}}^{ - 2\sigma }}/s{{\left( {{q^3} + {r^3}} \right)}^2}} |$ and $| { - {e^{ - 2\sigma }}} |$ in the equation of motion for $r$, as well as $|3{q^6}{{\rm{e}}^{ - 2\sigma }}\left( {{q^3} - 2{r^3}} \right)/s{\left( {{q^3} + {r^3}} \right)^3}|$ in the equation of motion for $\sigma$ become negligible as $r $ approaches zero.
The near-singularity dynamics then take the simplified form
\begin{equation}\label{N001}
r{r_{,tt}} \approx  - r_{,t}^2,
\end{equation}
\begin{equation}\label{N001}
{\sigma _{,tt}} \approx \varphi _{,t}^2,
\end{equation}
\begin{equation}\label{N001}
{\varphi _{,tt}} \approx  - 2\frac{{{r_{,t}}{\varphi _{,t}}}}{r}.
\end{equation}
This dynamics exhibits striking similarity to collapse of a neutral, strong scalar field leading to the formation of a Schwarzschild black hole~\cite{Gnedin:1993nau,Brady:1995ni,Guo:2013dha,Guo:2015laa}, which implies that the singularity in our model is spacelike. 

\subsection{Collapse of a critical scalar field in Hayward geometry}
\begin{figure*}[!]
\centering
\includegraphics[scale=0.395]{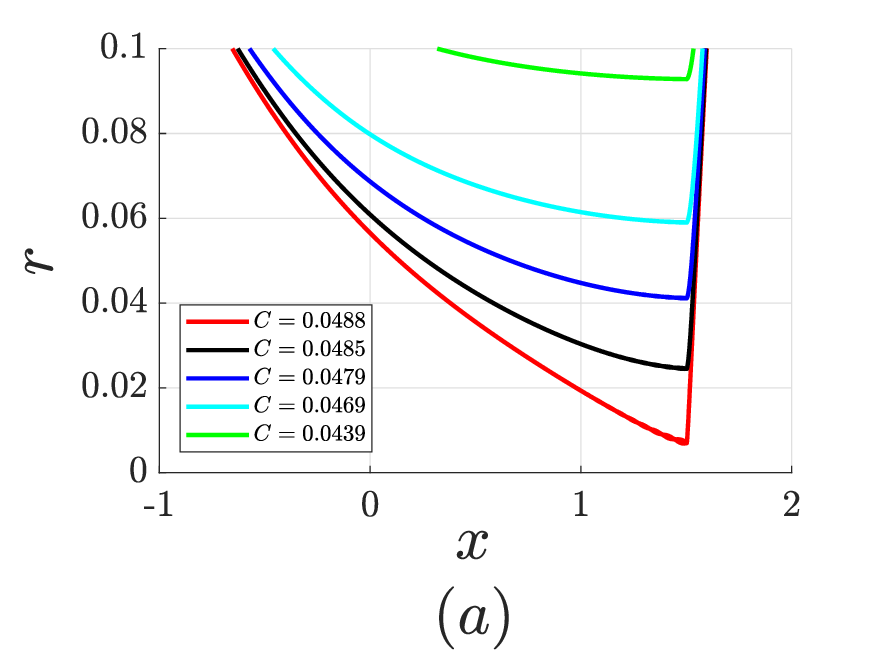}
\includegraphics[scale=0.395]{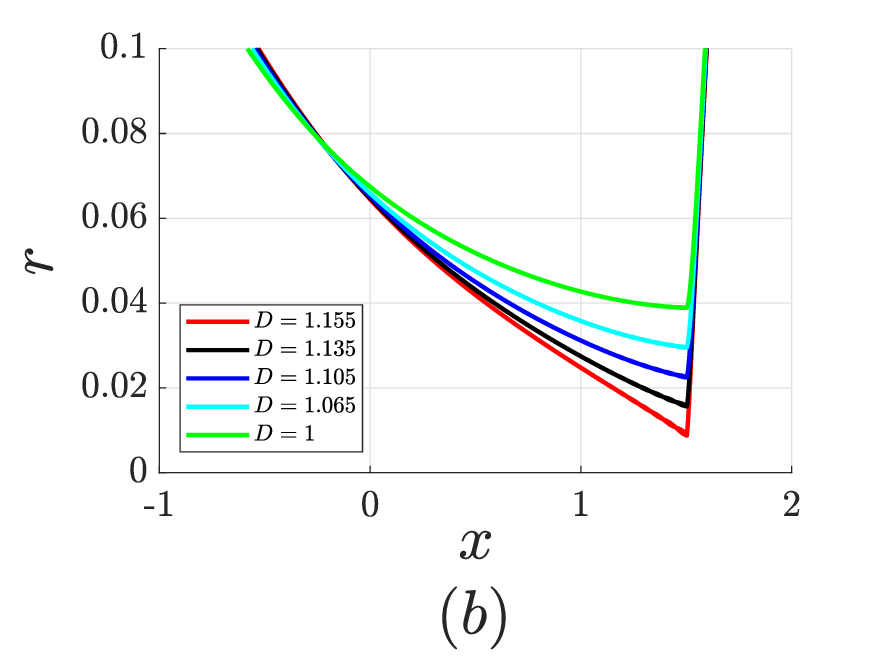}
\includegraphics[scale=0.395]{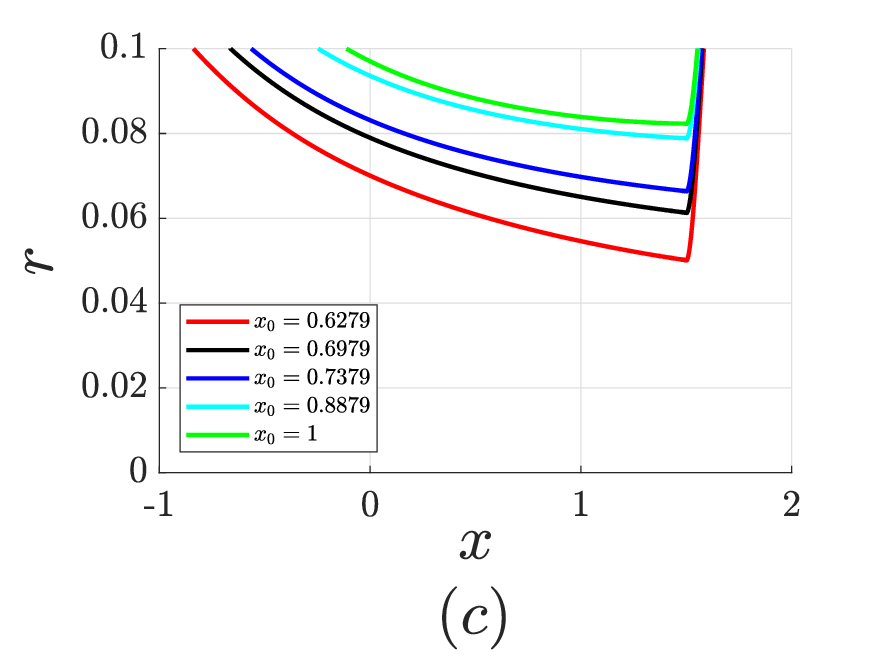}
\caption{Final state of \( r \) for $\varphi (0,x) = C\exp [ - {(x - {x_0})^2}/D]$:
(a): varying amplitude \( C \) with \( D = 1 \) and \( x_0 = 1 \);
(b): varying width \( D \) with \( C = 0.048 \) and \( x_0 = 1 \);
(c): varying central position \( x_0 \) with \( C = 0.048 \) and \( D = 1 \).
Increasing $C$ and ${D}$, or decreasing $x_0$, leads to a contraction of the inner horizon.
Once the critical values are reached, the inner horizon disappears, forming the $r=0$ singularity.
}\label{Fig7}
\end{figure*}

\begin{figure*}[!]
\centering
\includegraphics[scale=0.395]{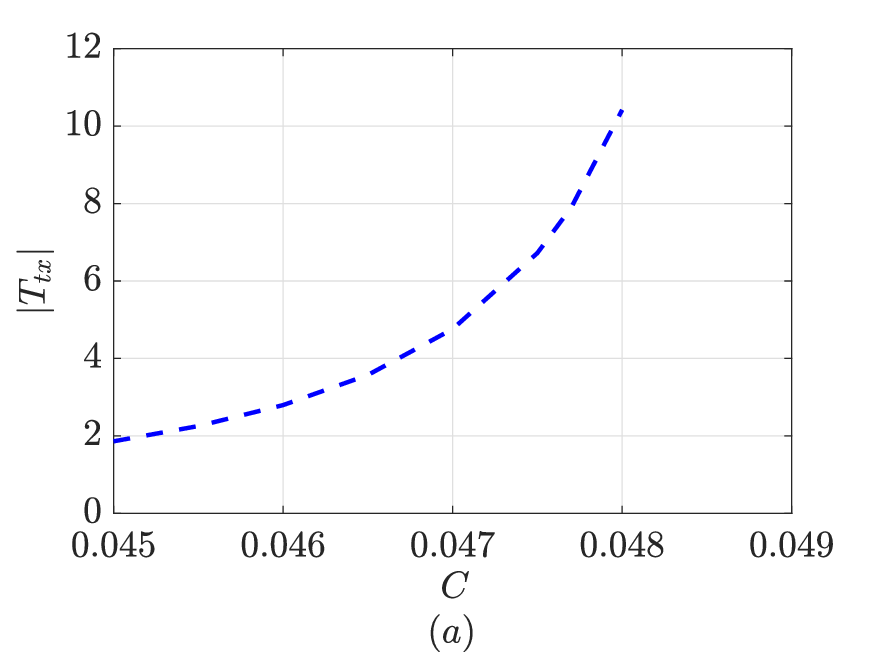}
\includegraphics[scale=0.395]{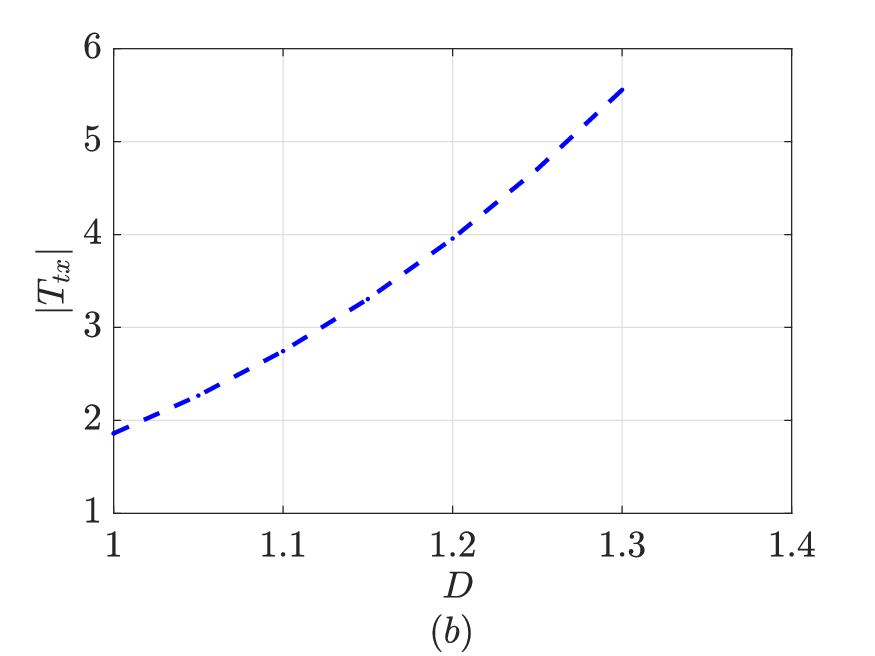}
\includegraphics[scale=0.395]{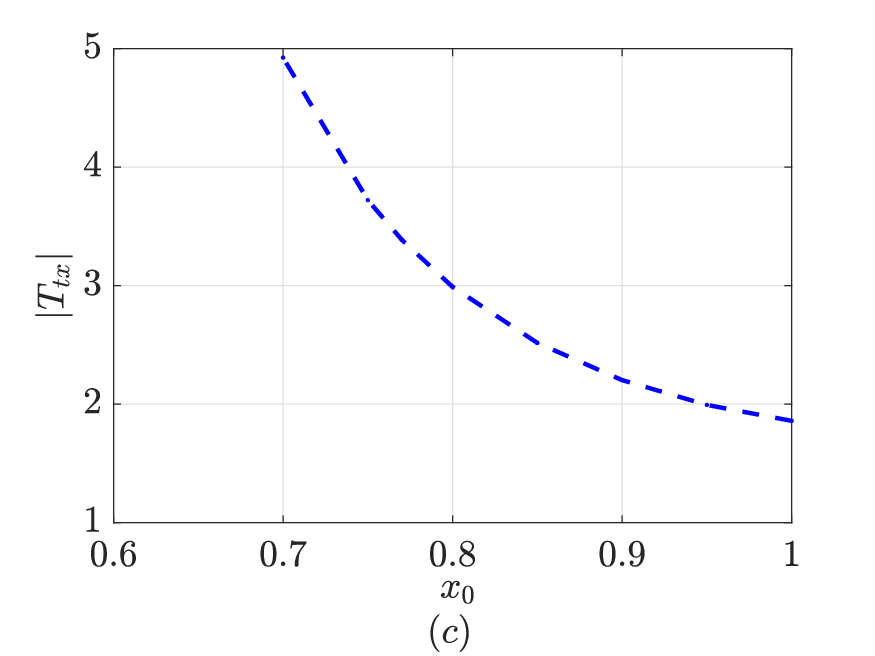}
\caption{Magnitude of the time-space component of the scalar-field energy-momentum tensor, $\left| {{T_{tx}}} \right|$, for $\varphi (0,x) = C\exp [ - {(x - {x_0})^2}/D]$.
The quantity is evaluated on the slice $x=0.5$, at a reference stage when the scalar pulse has propagated to the inner-horizon region and the contraction of the inner horizon has become apparent.
(a): varying amplitude \( C \) with \( D = 1 \) and \( x_0 = 1 \);
(b): varying width \( D \) with \( C = 0.045 \) and \( x_0 = 1 \);
(c): varying central position \( x_0 \) with \( C = 0.045 \) and \( D = 1 \).
}\label{Fig8}
\end{figure*}

Based on our preceding analysis, we observe that the areal radius $r$ is sensitive not only to the amplitude of the scalar field but also to its characteristic width and central position, all of which influence the contraction of the inner horizon of the black hole, as illustrated in Fig.~\ref{Fig7}.
In the final state of the evolution, $r$ decreases as the parameters $C$ and $D$ are increased, or as $x_0$ is decreased.
Once the critical values are reached, an $r=0$ singularity forms.
While $C$ directly controls the amplitude of the perturbation, the parameters $D$ and $x_0$ influence the dynamics more subtly by shaping the scalar flux profile and shifting the arrival time of the energy flux propagating into the black-hole interior.
To further illustrate the role of energy transport in this process, we evaluate the time-space component of the scalar-field energy-momentum tensor ${T_{tx}} = {\varphi _{,t}}{\varphi _{,x}}$, for $\varphi (0,x) = C\exp [ - {(x - {x_0})^2}/D]$ corresponding to Fig.~\ref{Fig2}(b).
As shown in Fig.~\ref{Fig8}, we examine the magnitude $\left| {{T_{tx}}} \right|$ on the slice $x=0.5$, at a reference stage when the scalar pulse has propagated to the inner-horizon region and the contraction of the inner horizon has become apparent, while varying $C$, $D$ and $x_0$ separately.
It is observed that $\left| {{T_{tx}}} \right|$ increases with larger values of $C$ and $D$, and with smaller values of $x_0$.
This trend mirrors the behavior of inner-horizon contraction depicted in Fig.~\ref{Fig7}.
Such correspondence is consistent with the view that scalar energy transport plays an important role in shaping the parameter dependence of the interior dynamics.
This illustration is intended to reveal the underlying physical mechanism. 
Moreover, the contraction of the inner horizon is not solely dictated by a single flux, but emerges from the nonlinear backreaction between the scalar field and the spacetime geometry.
Together, these perturbation parameters modify the scalar energy transport into the black-hole interior and its temporal distribution, which in turn affect the subsequent nonlinear evolution associated with mass inflation and inner-horizon contraction.

\begin{figure*}[!]
\centering
\includegraphics[scale=0.395]{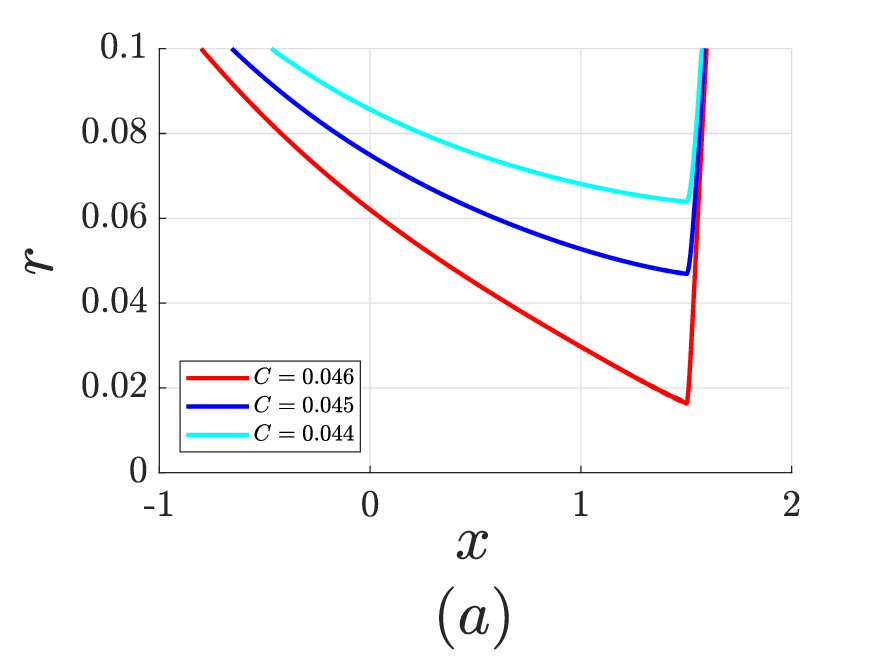}
\includegraphics[scale=0.395]{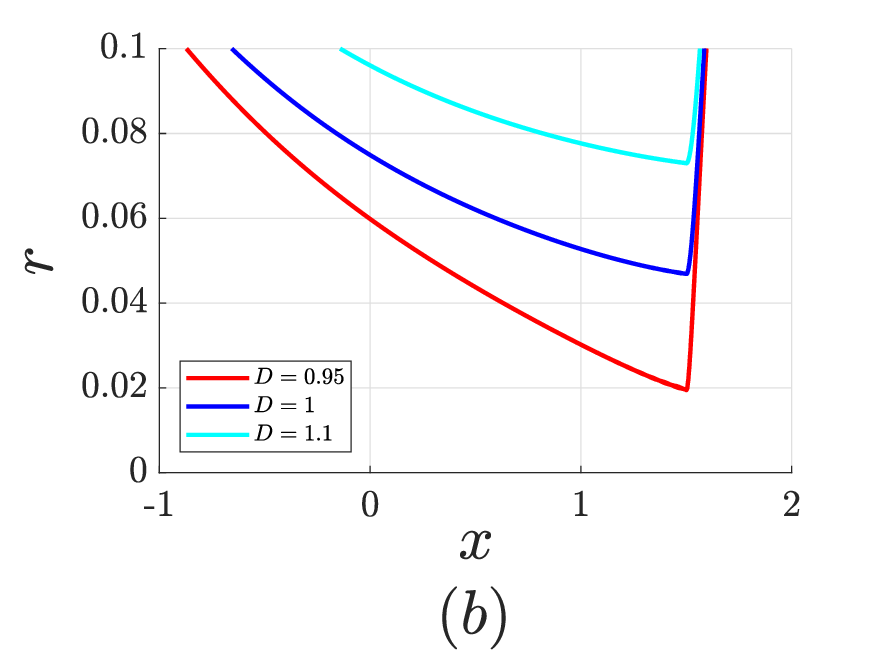}
\includegraphics[scale=0.395]{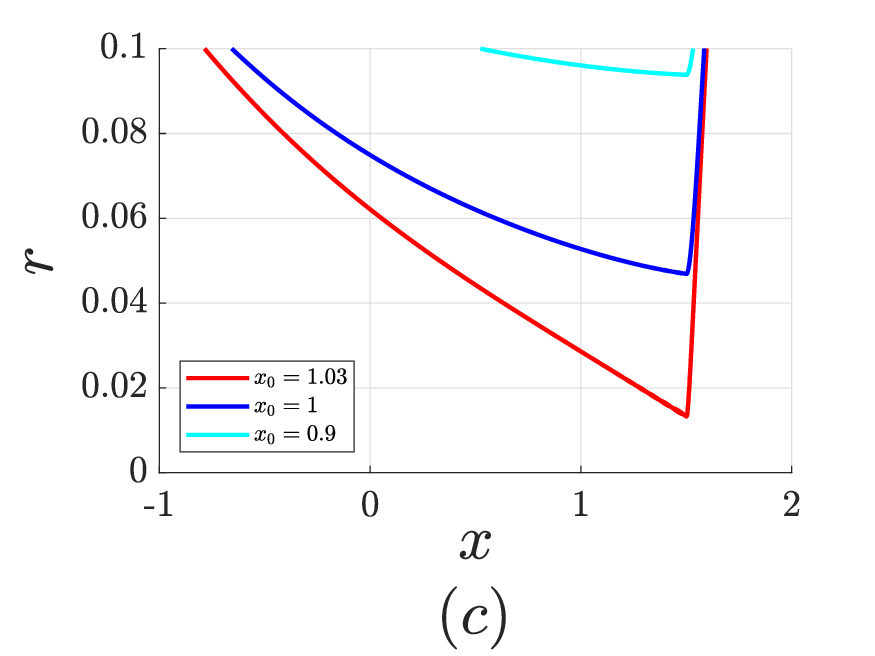}
\caption{Final state of \( r \) for $\varphi (0,x) = C\tanh [(x - {x_0})/D]$.
(a): varying amplitude \( C \) with \( D = 1 \) and \( x_0 = 1 \);
(b): varying width \( D \) with \( C = 0.045 \) and \( x_0 = 1 \);
(c): varying central position \( x_0 \) with \( C = 0.045 \) and \( D = 1 \).
The inner horizon shrinks with increasing $C$ and ${x_0}$ or decreasing $D$.
Once the critical values are reached, the inner horizon disappears, forming the $r=0$ singularity.
}\label{Fig9}
\end{figure*}

\begin{figure*}[!]
\centering
\includegraphics[scale=0.5]{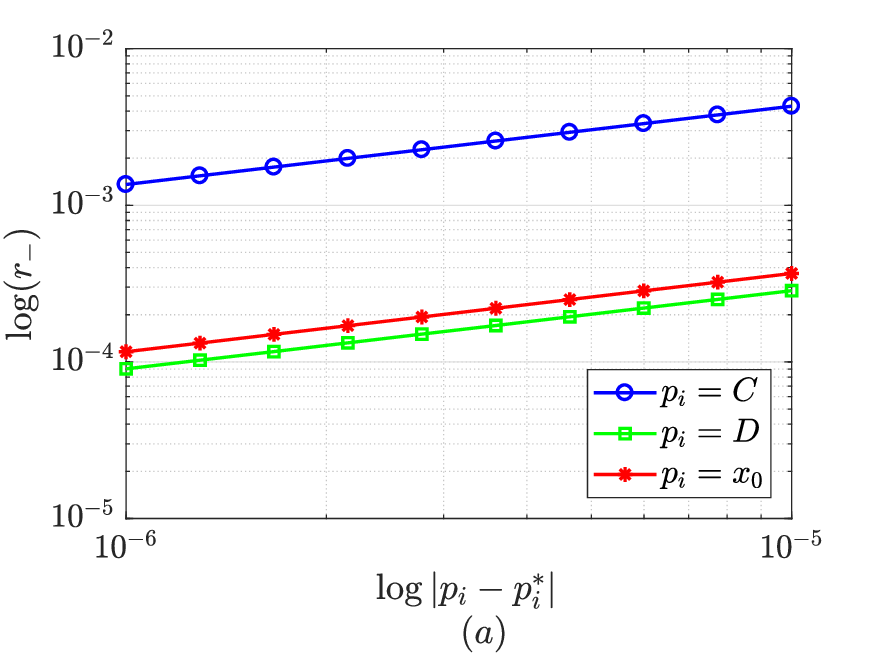}
\includegraphics[scale=0.5]{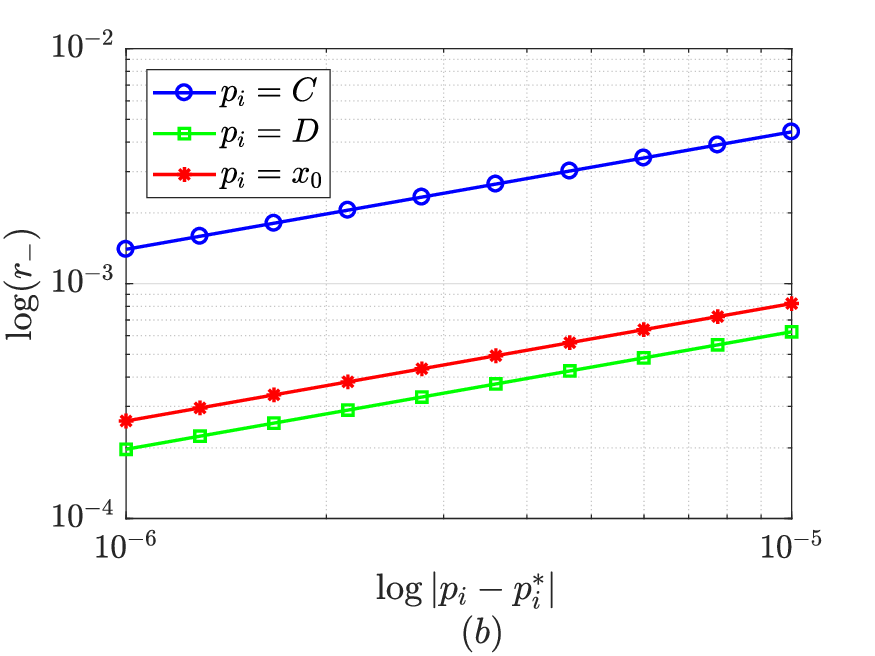}
\caption{ The radius of the inner horizon vs. the parameter $p_i$.
(a) Critical values ${C_*}= 0.048825647950$ (with $D=1$, ${x_0}= 1$), ${D_*}= 1.164342471646$ (with $C=0.048$, ${x_0}= 1$), and ${x_0*}= 1.243714430254$ (with $C=0.048$, ${D}= 1$) for the scalar-field profile $\varphi (0,x) = C\exp [ - {(x - {x_0})^2}/D]$.
(b) Critical values ${C_*}= 0.046135786121$ (with $D=1$, ${x_0}= 1$), ${D_*}= 0.940146912036$ (with $C=0.045$, ${x_0}= 1$), and ${x_0*}= 1.032592785530$ (with $C=0.045$, ${D}= 1$) for the scalar-field profile $\varphi (0,x) = C\tanh [(x - {x_0})/D]$.
A clear power-law scaling is observed between $\log ({r_ - })$ and $\log |p - {p_*}|$, such that ${r_ - } \propto |p - {p_*}{|^\gamma }$, where the critical exponent $ \gamma $ is approximately 0.5.
}\label{Fig10}
\end{figure*}

This behavior motivates a systematic investigation of the functional dependence, enabling a quantitative characterization of the underlying nonlinear dynamics in the vicinity of criticality.
Here we focus on the physically relevant scenario of scalar-field collapse near the critical threshold, using the Gaussian initial profile as a representative case.
For initial data parameterized by $p$ (which could be $C$, $D$, or ${x_0}$ in this case), there exists a critical threshold value ${p_*}$.
For \( p > p^* \), the inner horizon $r$ vanishes, whereas for \( p < p^*\), it converges to a finite, non-zero value.
For values of $p$ sufficiently close to ${{p_*}}$, we have found the relationship can be approximated as
\begin{equation}\label{N001}
{r_ - } \propto {\left| {p - {p_*}} \right|^\gamma }. 
\end{equation}
The power-law scaling is shown in Fig.~\ref{Fig10}(a), where ${\rm{log}}|p-{p_*}|$ is plotted against ${\rm{log}}({r_-})$ with critical values: ${C_*}= 0.048825647950$ (field amplitude, with $D=1$, ${x_0}= 1$), ${D_*}= 1.164342471646$ (width parameter, with $C=0.048$, ${x_0}= 1$), and ${x_{0*}}= 1.243714430254$ (peak position, with $C=0.048$, ${D}= 1$) for the initial profile.
The critical exponent $\gamma $ is $0.499975 \pm 0.000002$ for $C$, $0.4995 \pm 0.0002$ for $D$ and $0.50013 \pm 0.00006$ for ${x_0}$, as determined through numerical fitting.
This result naturally extends to the broader class of Reissner-Nordström black holes, where we demonstrate that the observed relationship persists~\cite{Shao:2025fki}.

As a robustness check, we also consider alternative smooth, localized initial profiles beyond the Gaussian packet, such as a tanh-type pulse:
\begin{equation}\label{N001}
\varphi (0,x) = C\tanh (\frac{{x - {x_0}}}{D}).
\end{equation}
We find that the qualitative dependence of the inner-horizon contraction on the initial-data parameters persists, as shown in Fig.~\ref{Fig9}.
This demonstrates that the observed inner-horizon contraction is not artifacts of the specific Gaussian choice, but rather reflect a more general response of the Hayward geometry to localized scalar-field perturbations.
Moreover, as shown in Fig.~\ref{Fig10}(b), under tanh-type pulse perturbations, we observe power-law scaling near criticality, further reinforcing the universality of this phenomenon.

\section{\label{section4}Concluding Remarks}
In this study, we have explored collapse dynamics of a neutral scalar field in Hayward black hole, examining how the scalar field influences the inner horizon and the overall structure of the black hole. 
We considered various field strengths, ranging from weak to strong perturbations, and identified distinct behaviors corresponding to these regimes. 
For weak scalar perturbations, the inner horizon of the Hayward black hole retains a finite radius, while strong scalar fields induce significant changes, leading to the contraction of the inner horizon and the potential formation of a spacelike singularity.
We observed that the behavior of the inner horizon is strongly affected not only by the amplitude of the scalar field but also by its width and central location.
In particular, in the near-critical regime where characteristic parameter of a scalar field approaches a critical threshold, we found that the radius of the inner horizon exhibits a scaling behavior \(r_- \propto |p - p_*|^\gamma\), with a critical exponent \(\gamma \approx 0.5\), revealing rich critical behavior in the black hole's internal structure. 
Moreover, the contraction dynamics of the inner horizon and the associated mass inflation phenomena highlight the role of scalar field perturbations in driving the evolution of regular black holes towards a possible singularity. 
These findings not only shed light on the dynamics of black hole interiors, but also provide a novel testing ground for potential gravitational-wave observations.
In future work, it would be interesting to explore the effects of other types of matter fields and the influence of higher-dimensional black holes on these dynamics. 

\section*{Acknowledgements}
The authors thank Yongqiang Wang for helpful discussion.
This work is supported in part by the National Key R\&D Program of China, Grant No. 2020YFC2201300, No. 2021YFC2203001, and the National Natural Science Foundation of China, Grants No. 12035016, No. 12375058,  No. 12361141825, No. 12447182, and No. 12575047.

\bibliography{mybibfile}

\end{document}